\documentclass{elsart}
\usepackage{latexsym, amssymb, enumerate, amsmath}
\usepackage {graphicx}

\begin{document}

\begin{frontmatter}
\centering{ Physica A 389 (2010) 3193--3217}

\title{\textbf{Correlations, Risk and Crisis:\\ From Physiology to Finance}}

\author{Alexander N. Gorban\corauthref{cor1}}
\ead{ag153@le.ac.uk}
\corauth[cor1]{Corresponding author: Centre
for Mathematical Modelling, University of Leicester, University
Road, Leicester, LE1 7RH,  UK}
\address{University of Leicester,
Leicester,  LE1 7RH,   UK}

\author{Elena V. Smirnova}
\ead{seleval2008@yandex.ru}
\address{Siberian Federal University, Krasnoyarsk, 660041, Russia}

\author{Tatiana A. Tyukina}
\ead{tt51@le.ac.uk}
\address{University of Leicester,
Leicester,  LE1 7RH,  UK}

\date{}

\maketitle

\begin{abstract}
We study the dynamics of correlation and variance in systems under
the load of environmental factors. A universal effect in ensembles
of similar systems under the load of similar factors is described:
in crisis, typically, even before obvious symptoms of crisis
appear, correlation increases, and, at the same time, variance
(and volatility) increases too. This effect is supported by many
experiments and observations of groups of humans, mice, trees,
grassy plants, and on financial time series.

A general approach to the explanation of the effect through
dynamics of individual adaptation of similar non-interactive
individuals to a similar system of external factors is developed.
Qualitatively, this approach follows Selye's idea about adaptation
energy.
\end{abstract}

\end{frontmatter}

\section*{Introduction: Sources of Ideas and Data}

In many areas of practice, from physiology to economics,
psychology, and engineering we have to analyze the behavior of
groups of many similar systems, which are adapting to the same or
similar environment. Groups of humans in hard living conditions
(Far North city, polar expedition, or a hospital, for example),
trees under the influence of anthropogenic air pollution, rats
under poisoning, banks in financial crisis, enterprises in
recession, and many other situations of that type provide us with
plenty of important problems, problems of diagnostics and
prediction.

\begin{figure} \centering{
\includegraphics[width=100mm]{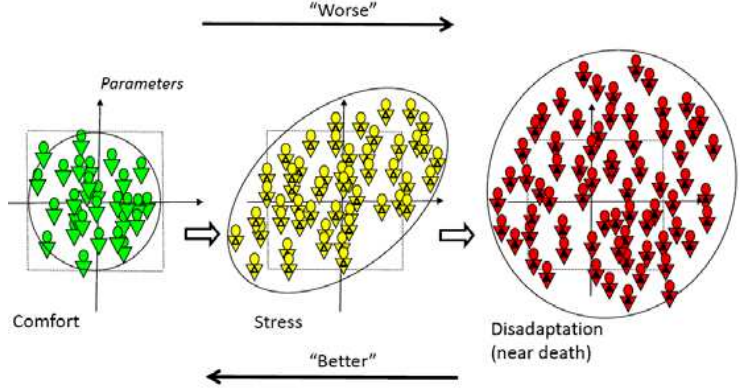}}
\caption{\label{Fig:effect} Correlations and variance in crisis.
The typical picture: $Cor \uparrow ; \, Var \uparrow$ -- stress;
$Cor \downarrow ; \, Var \downarrow$ -- recovering; $Cor
\downarrow ; \, Var \uparrow$ -- approaching the disadaptation
catastrophe after the bottom of the crisis. In this schematic
picture, axes correspond to attributes, normalized to the unite
variance in the comfort state.}
\end{figure}

For many such situations it was found that the correlations
between individual systems are better indicators than the value of
attributes. More specifically, in thousands of experiments it was
shown that in crisis, typically, even before obvious symptoms of
crisis appear, the correlations increase, and, at the same time,
the variance (volatility) increases too (Fig.~\ref{Fig:effect}).

On the other hand, situations with inverse behavior were
predicted theoretically and found experimentally
\cite{Mansurov}. For some systems it was demonstrated that
after the crisis achieves its bottom, it can develop into two
directions: recovering (both the correlations and the variance
decrease) or fatal catastrophe (the correlations decrease, but
the variance continues to increase) (Fig.~\ref{Fig:effect}).
This makes the problem more intriguing.

If we look only on the state but not on the history then the
only difference between comfort and disadaptation in this
scheme is the value of variance: in the disadaptation state the
variance is larger and the correlations in both cases are low.
Qualitatively, the typical behavior of an ensemble of similar
systems, which are adapting to the same or similar environment
looks as follows:
\begin{itemize}
\item{In a well-adapted state  the deviations of the systems' state
from the average value have relatively low correlations;}
\item{Under increasing of the load of environmental factors some
of the systems leave the low-correlated comfort cloud and form a
low-dimensional highly correlated group (an {\em order parameter}
appears). With further increasing of the load more systems join
this highly correlated group. A simplest model based on the
Selye's ideas about adaptation gives the explanation of this
effect (see Sec.\ref{Selye Model});}
\item{After the load gets over some critical value, the
    order parameter disappears and the correlations
    decrease but the variance continues to increase.}
\end{itemize}
There is no proof that this is the only scenario of the changes.
Perhaps, it is not. It depends on the choice of parameters, for
example. Nevertheless, the first part (appearance of an order
parameter) was supported by plenty of experiments and the second
part (destroying of the order parameter) is also supported by
observation of the systems near death.

Now, after 21 years of studying of this effect
\cite{GorSmiCorAd1st,Sedov},  we maintain that it is universal for
groups of similar systems that are sustaining a stress and have an
adaptation ability. Hence, a theory at an  adequate level of
universality is needed.

In this paper we review some data for different kinds of systems:
from humans to plants
\cite{Sedov,Pokidysheva,Svetlichnaia,RazzhevaikinObese2007,mice,RazzhevaikinTrava1996},
and perform also a case study of the thirty largest companies from
the British stock market for the period 2006--2008.

In economics, we use also published results of data analysis for
equity markets of seven major countries over the period 1960--1990
\cite{LonginCorrNonconst1995}, for the twelve largest European
equity markets after the 1987 international equity market crash
\cite{MericEuMarket1987}, and for thirty companies from Deutsche
Aktienindex (DAX) over the period 1988-1999
\cite{DrozdComCollNoise2000}.The analysis of correlations is very
important for portfolio optimization, and an increase of
correlations in a crisis decreases the possibility of risk
diversification (\cite{Stanley2000}, Chs. 12, 13). In 1999, it was
proposed \cite{Mantegna1999} to use the distance $d_{ij}=
\sqrt{2(1-\rho _{ij})}$, where $\rho _{ij}$ is the correlation
coefficient, for the analysis of the hierarchical structure of a
market. (This distance for multidimensional time series analysis
was analyzed previously in Ref. \cite{Gower1966}.) The performance
of this approach was demonstrated on the stocks used to compute
the Dow Jones Industrial Average and on the portfolio of stocks
used to compute the S\&P 500 index. This approach was further
developed and applied (together with more standard correlation
analysis) for analysis of anatomy of the Black Monday crisis
(October 19, 1987) \cite{ChakrabortiMarketCorr2003}. In this
analysis, hundreds of companies were used.

Stock price changes of the largest 1000 U.S. companies were
analyzed for the 2-year period 1994--1995 \cite{Stanley1999}, and
statistics of several of the largest eigenvalues of the
correlation matrix were evidenced to be far from the random matrix
prediction. This kind of analysis was continued for the three
major US stock exchanges, namely, the New York Stock Exchange
(NYSE), the American Stock Exchange (AMEX), and the National
Association of Securities Dealers Automated Quotation (NASDAQ)
\cite{Stanley2002}. Cleaning the correlation matrix by removing
the part of the spectrum explainable by random matrix ensembles
was proposed \cite{Potters2005}. Spectral properties of the
correlation matrix were analyzed also for 206 stocks traded in
Istanbul Stock Exchange Market during the 5-year period 2000--2005
\cite{SadikCrosCorFin2007}.

Linear and nonlinear co-movements presented in the Real Exchange
Rate (RER) in a group of 28 developed and developing countries were
studied to clarify the important question about ``crisis contagion"
\cite{MatesanzDependenceCriz2008}: Do strong correlations appear
before crisis and provide crisis contagion, or do they grow stronger
because of crisis? The spread of the credit crisis (2007--2008) was
studied by referring to a correlation network of stocks in the S\&P
500 and NASDAQ-100 indices. Current trends demonstrate that the
losses in certain markets, follow a cascade or epidemic flow along
the correlations of various stocks. But whether or not this idea of
epidemic or cascade is a metaphor or a causal model for this crisis
is not so obvious \cite{Smith2009}.

Most of the data, which we collected by ourselves or found in
publications, support the hypothesis presented in
Fig.~\ref{Fig:effect}. In all situations, the definitions of
stress and crisis were constructed by experts in specific
disciplines on the basis of specific knowledge. What do ``better"
and ``worse" mean? This is a nontrivial special question and from
the point of view of very practically oriented researchers the
main outcome of modeling may be in the  definition of crisis
rather than in the explanation of details
\cite{EliassonCrizDefin2001}. In many situations we can detect
that one man's crisis is another man's road to prosperity.

Nevertheless, all the experiments are unbiased in the following
sense: the definitions of the ``better--worse" scale were done
before the correlation analysis and did not depend on the results
of that analysis. Hence, one can state, that the expert evaluation
of the stress and crisis can be (typically) reproduced by the
formal analysis of correlations and variance.

The basic model of such generality should include little detail, and
we try to make it as simple as possible. We represent the systems,
which are adapting to stress, as the systems which optimize
distribution of available amount of resource for the neutralization
of different harmful  factors (we also consider deficit of anything
needful as a harmful factor).

The crucial question for these {\it factor--resource} models is:
what is the {\it resource of adaptation}? This question arose for
the first time when Selye published the concept of {\it adaptation
energy} and experimental evidence supporting this idea
\cite{SelyeAEN,SelyeAE1}. After that, this notion was
significantly improved \cite{GP_AE1952}, plenty of indirect
evidence supporting this concept was found, but this elusive
adaptation energy is still a theoretical concept, and in the
modern ``Encyclopedia of Stress" we read: ``As for adaptation
energy, Selye was never able to measure it..." \cite{AEencicl}.
Nevertheless, the notion of adaptation energy is very useful in
the analysis of adaptation and is now in wide use (see, for
example, \cite{BreznitzAEappl,SchkadeOccAdAE2003}).

The question about the nature of adaptation resource remains
important for the economic situation too. The idea of {\it
exchange} helps here: any resource could be exchanged for another
one, and the only question is -- what is the ``exchange rate", how
fast this exchange could be done, what is the margin, how the
margin depends on the exchange time, and what is the limit of that
exchange. In the zero approximation we can just postulate the
universal adaptation resource and hide all the exchange and
recovering processes. For biophysics, this exchange idea seems
also attractive, but of course there exist some limits on the
possible exchange of different resources. Nevertheless, we can
follow the Selye arguments and postulate the adaptation energy
under the assumption that this is not an ``energy", but just a
pool of various exchangeable resources. When an organism achieves
the limits of resources exchangeability, the universal
non-specific stress and adaptation syndrome transforms
(disintegrates) into specific diseases. Near this limit we have to
expect the {\it critical retardation} of exchange processes.

In biophysics, the idea of optimization requires additional
explanation. The main source of the optimality idea in biology is
the formalization of natural selection and adaptive dynamics.
After works of Haldane (1932) \cite{Haldane} and Gause (1934)
\cite{Gause} this direction, with various concepts of {\it
fitness} optimization, was further developed (see, for example,
review papers \cite{Bom02,Oechssler02,GorbanSelTth}). To transfer
the evolutionary optimality principles to short and long term
adaptation we need the idea of {\it genocopy-phenocopy
interchangeability} (\cite{West-Eberhardgenocopy-phenocopy}, p.
117). The phenotype modifications simulate the optimal genotype,
but in a narrower interval of change. We can expect that
adaptation also gives the optimal phenotype, but the interval of
the possible changes should be even narrower, than for
modifications. The idea of convergence of genetic and
environmental effects was supported by analysis of genome
regulation \cite{ZuckerkandlConvergGenEnv} (the principle of
concentration-affinity equivalence). This gives a basis for the
optimality assumption in adaptation modeling. For ensembles of
man-made systems in economics, the idea of optimality also can be
motivated by selection of strategies arguments.

To analyze resource redistribution for the compensation of
different environmental factors we have to answer one more
question: how is the system of factors organized? Ecology already
has a very attractive version for an answer. This is Liebig's Law
of the Minimum. The principle behind this law is quite simple.
Originally, it meant that the scarcest necessity an organism
requires will be the limiting factor to its performance. A bit
more generally, the worst factor determines the situation for an
organism, and free resource should, perhaps, be assigned for
neutralization of that factor (until it loses its leadership).

The opposite principle of factor organization is synergy: the
superlinear mutual amplification of factors. Adaptation to
Liebig's system of factors, or to any synergistic system, leads to
two paradoxes of adaptation:
\begin{itemize}
\item{{\it Law of the Minimum paradox} (Sec.~\ref{Sec:LawMinParad}):
If for a randomly selected
pair, ("State of environment -- State of organism"), the Law of the
Minimum is valid (everything is limited by the factor with the worst
value) then, after adaptation, many factors (the maximally possible
amount of them) are equally important. }
\item{{\it Law of the Minimum inverse paradox} (Sec.~\ref{Sec:LawMinInvPar}): If for a randomly
selected pair, ("State of environment -- State of organism"), many
factors are equally important  and superlinearly amplify each
other then, after adaptation, a smaller amount of factors is
important (everything is limited by the factors with the worst
non-compensated values, the system approaches the Law of the
Minimum).}
\end{itemize}

After introduction of the main ideas and data sources, we are in a
position to start more formal consideration.

\section{Indicators}

How can we measure correlations between various attributes in a
population? If we have two variables, $x$ and $y$,  the answer is
simple: we measure $(x_i, y_i)$ for different individuals ($i=1,...
n$, $n>1$ is the number of measurements). The sample correlation
coefficient (the Pearson coefficient) is
\begin{equation}\label{CoCor}
r = \frac{\langle x y \rangle - \langle x\rangle \langle y
\rangle}{\sqrt{\langle(x_i -\langle x \rangle)^2\rangle}
\sqrt{\langle(y_i -\langle y \rangle)^2\rangle}}
\end{equation}
where $\langle ... \rangle $ stands for the sample average value:
$\langle x \rangle =\frac{1}{n}\sum_i x_i$.

If individuals are characterized by more than two attributes
$\{x^l | l=1, ... m\}$ then we have $m(m-1)/2$ correlation
coefficients between them, $r_{jk}$. In biophysics, we usually
analyze correlations between attributes, and each individual
organism is represented as a vector of attribute values.

In analysis of financial time series, the standard situation
may be considered as a ``transposed" one. Each object (stock,
enterprise, ...) is represented by a vector of values of a
variable (asset return, for example) in a window of time and we
study correlations between objects. This is, essentially, just
a difference between $X$ and $X^T$, where $X$ is the matrix of
data. In correlation analysis, this difference appears in two
operations: centralization (when we subtract means in the
computation of covariance) and normalization (when we transform
the covariance into the correlation coefficient). In one case,
we centralize and normalize the columns of $X$: subtract
average values in columns, and divide columns on their standard
deviations. In another case, we apply these operations to the
rows of $X$. For financial time series, the synchronous
averages and variances (``varieties") and time averages and
variances (``volatilities") have different statistical
properties. This was clearly demonstrated in a special case
study \cite{VarVolLillo2000}.

Nevertheless, such a difference does not appear very important for
the analysis of the total level of correlations in crisis (just
the magnitude of correlation changes, and correlations in time are
uniformly less than synchronous ones, this is in agreement with
observations from Ref.~\cite{VarVolLillo2000}). More details are
presented in the special case study below.

In our case study we demonstrated that in the analysis of
financial time series it may be also convenient to study
correlations between parameters, not between individuals. It means
that we can study correlation between any two time moments and
consider data from different individuals as values of random 2D
vector. It is necessary to stress that this correlations between
two time moments are very different from the standard
autocorrelations for stationary time series (which characterize
the sample of all pairs of time moments with a given lag in time).

For example, let $X_{it}$ be a log-return value for $i$th stock at
time moment $t$ ($i=1,...n$, $t=\tau + 1,...\tau+T$). Each row of
the data matrix $X_{it}$ corresponds to an individual stock and
each column corresponds to a time moment. If we normalize and
centralize data in rows and calculate the correlation coefficients
between rows ($r_{ij}=\sum_t X_{it}X_{jt}$ for centralized and
normalized data) then we find the correlations between stocks. If
we normalize and centralize data in columns and calculate the
correlation coefficients between them ($r_{t_1 t_2}=\sum_i
X_{it_1}X_{it_2}$ for centralized and normalized data) then we
find the correlations between time moments. In crisis, dynamics of
the correlations between stocks is similar to behavior of the
correlations  between time moments. One benefit from use of the
correlations between time moments is absence of averaging in time
(locality): this correlation coefficient depends on data at two
time moments. This allows to analyze the anatomy of crisis in
time.

To collect information about correlations between many attributes
in one indicator, it is possible to use various approaches. Fist
of all, we can evaluate the non-diagonal part of the correlation
matrix in any norm, for example, in $L_p$ norm
\begin{equation}\label{lpnorm}
\| r \|_p = \left(\sum_{j>k} |r_{jk}|^p\right)^{\frac{1}{p}}.
\end{equation}
If one would like to study strong correlations, then it may be
better to delete terms with values below a threshold $\alpha>0$ from
this sum:
\begin{equation}\label{lpweight}
 G _{p, \alpha} = \left(\sum_{j>k, \ |r_{jk}|> \alpha}
|r_{jk}|^p\right)^{\frac{1}{p}}.
\end{equation}
This quantity $ G _{p, \alpha}$ is a $p$-weight of the $\alpha$-{\it
correlation graph}. The vertices of this graph correspond to
variables, and these vertices are connected by edges, if the
absolute value of the correspondent sample correlation coefficient
exceeds $\alpha$: $|r_{jk}|>\alpha$. In practice, the most used
indicator is the weight $G= G _{1, 0.5}$, which corresponds to $p=1$
and $\alpha=0.5$.

The correlation graphs are used during decades for visualization
and analysis of correlations (see, for example,
\cite{GorSmiCorAd1st,Whittaker1990,Brillinger1996}). Recently, the
applications of this approach is intensively developing in data
mining \cite{Fried2003,Verma2005,Huynh2006} and econophysics
\cite{Onella1,Onella2}.

Another group of indicators is produced from the principal
components of the data. The principal components are eigenvectors of
the covariance matrix and depend on the scales. Under normalization
of scales to unit variance, we deal with the correlation matrix. Let
$\lambda_1 \geq \lambda_2 \geq ... \lambda_m \geq 0$ be eigenvalues
of the correlation matrix. In this paper, we use the eigenvalues and
eigenvectors  of the correlation matrix. It is obvious that
$\langle\lambda \rangle =1$ and $m^{p-1}\geq \langle\lambda^p
\rangle \geq 1$ for $p>1$, $ \langle\lambda^p \rangle = 1$ if all
non-diagonal correlation coefficients are zero and $
\langle\lambda^p \rangle = m^{p-1}$ if all correlation coefficients
are $\pm 1$. To select the {\it dominant} part of principal
components, it is necessary to separate the ``random" part from the
``non-random" part of them. This separation was widely discussed
(see, for example, the expository review
\cite{CangelosiBrockStick}).

The simplest decision gives {\it Kaiser's significance rule}: the
significant eigenvalues are those, which are greater than the
average value: $\lambda_i>\langle\lambda \rangle$. For the
eigenvalues of the correlation matrix which we study here, it
means $\lambda_i>1$. This rule works reasonably well, when there
are several eigenvalues significantly greater than one and the
others are smaller, but for a matrix which is close to a random
matrix the performance may not be so good. In such cases this
method overestimates the number of principal components.

In econophysics, another simple criterion for selection of
dominant eigenvalues has become popular
\cite{Sengupta1999,Stanley2002,Potters2005,SadikCrosCorFin2007}.
Let us imagine that the dimension of the data vector $m$ is large
and the amount of data points $n$ is also large, but their ratio
$q=n/m$ is not. This is the typical situation when we analyze data
about thousands of stocks: in this case the time window could not
be much larger than the dimension of data vector. Let us compare
our analysis of real correlations to the fictitious correlations,
which appear in $m \times n$ data matrices with independent,
centralized, normalized and Gaussian matrix elements. The
distribution of the sample covariance matrix is the Wishart
distribution \cite{Wishart}. If $n \to \infty$ for given $m$ then
those fictitious correlations disappear, but if both $m, n \to
\infty$ for constant $q>1$ then there exists the limit
distribution of eigenvalues $\lambda$ with density
\begin{equation}\label{RMTeiganvalues}
\begin{split}
&\rho(\lambda)=\frac{q}{2\pi}\sqrt{\left(\frac{\lambda_{\max}}{\lambda}-1\right)
\left(1-\frac{\lambda_{\min}}{\lambda}\right)}\,; \ \ \lambda_{\min}
\leq {\lambda} \leq \lambda_{\max}; \\
 &\lambda_{\max / \min}= 1 + \frac{1}{q}\pm 2\sqrt{\frac{1}{q}} \ .
\end{split}
\end{equation}
If the amount of points is less than dimension of data, ($q<1$) the
same formula with substitution of $q$ by $1/q$ is valid for
distribution of non-zero eigenvalues.

Instead of Kaiser's rule for dominant eigenvalues of the correlation
matrix we get $\lambda_i>\lambda_{\max}$ with $\lambda_{\max}$ given
by Eq. (\ref{RMTeiganvalues}). If $q$ grows to $\infty$, this new
rule turns back into Kaiser's rule. If $q$ is minimal ($q=1$), then
the proposed change of Kaiser's rule is maximal, $\lambda_{\max}=4$
and for dominant eigenvalues of the correlation matrix it should be
$\lambda_i>4 $. This new estimate is just an analogue of Kaiser's
rule in the case when the amount of data vectors is compatible with
the dimension of the data space, and, therefore,  the data set is
far from the law of large numbers conditions.

Another popular criterion for the selection of dominant
eigenvalues gives the so-called {\it broken stick model}. Consider
the closed interval $J = [0,1]$. Suppose $J$ is partitioned into
$m$ subintervals by randomly selecting $m - 1$ points from a
uniform distribution in the same interval. Denote by $l_k$ the
length of the $k$-th subinterval in the descending order. Then the
expected value of $l_k$ is
\begin{equation}\label{BrockStickExp}
\mathbf{E}(l_k)=\frac{1}{m}\sum_{j=k}^m \frac{1}{j} \ .
\end{equation}
Following the broken stick model, we have to include into the
dominant part those eigenvalues $\lambda_k$ (principal components),
for which
\begin{equation}\label{BrockStickForm}
\frac{\lambda_1 }{\sum_i \lambda_i}>\mathbf{E}(l_1) \ \& \
\frac{\lambda_2}{\sum_i \lambda_i}
>\mathbf{E}(l_2)\ \& \ ... \ \& \ \frac{\lambda_k }{\sum_i \lambda_i}>\mathbf{E}(l_k) \ .
\end{equation}
If the amount of data vectors $n$ is less than the data
dimension $m$, then $m-n$ eigenvalues are zeros, and in
Eqs.~(\ref{BrockStickExp}), (\ref{BrockStickForm}) one should
take $n$ subintervals instead of $m$ ones.

It is worth mentioning that the trace of the correlation matrix is
$m$, and the broken stick model transforms  (for $m>n$) into
$\lambda_i
> \sum_{j=i}^m \frac{1}{j}$ ($i=1,... k$). From the practical point
of view, this method slightly underestimates the number of
dominant eigenvalues. There are other methods based on the random
matrices ensembles, but nobody knows the exact dimension of the
empirical data, and the broken stick model works satisfactorily
and remains ``the method of choice".

To compare the broken stick model to Kaiser's rule, let us mention
that the first principal component is always significant due to
Kaiser's rule (if there exists at least one nonzero non-diagonal
correlation coefficient), but in the broken stick model it needs
to be sufficiently large: the inequality $\lambda_1 > \sum_{j=1}^m
\frac{1}{j}$ should hold. In a high dimension $m$ we can
approximate the sum by the quadrature: $\lambda_1 \gtrsim \ln m$.

If we have the dominant eigenvalues, $\lambda_1 \geq  \lambda_2 \geq
... \lambda_l > 0$, $l < m$, then we can produce some other measures
of the sample correlation:
\begin{equation}
\frac{\lambda_1}{\lambda_{l}}; \;\; \sum_{j=1}^{l-1}
\frac{\lambda_j}{\lambda_{j+1}}; \;\; \frac{1}{m} \sum_{j=1}^{l}
\lambda_j\ .
\end{equation}
Together with $\langle\lambda^p \rangle$ ($p>1$, the usual choice
is $p=2$) this system of indicators can be used for an analysis of
empirical correlations.

Recently \cite{Heimo2008} eigenvalues and eigenvectors of the matrix
of the absolute values of the correlation coefficients were used for
analysis of the New York Stock Exchange (NYSE) traded stocks. The
transformation from the correlation matrix to the matrix of absolute
values was justified by interpreting the absolute values as measures
of interaction strength without considering whether the interaction
is positive or negative. This approach gives the possibility to
apply the classical theory of positive matrices as well as the
graphical representation of them.

The correlation matrix for financial time series is often positive.
Therefore, it is often possible to apply the theory of positive
matrices to analysis of correlations in financial time series.

The choice of possible indicators is very rich, but happily,
many case studies have shown that in analysis of adaptation the
simplest weight $G$ of the correlation graph performs well
(better or not worse than all other indicators -- see the case
study below). Similarity of behavior of various indicators,
from simple weight of the correlation graphs to more
sophisticated characteristics based on the principal component
analysis is expected. Nevertheless, it is desirable to
supplement the case studies by comparisons of behavior of
different indicators (for example, by scattering plots,
correlation analysis or other statistical tools). In our case
study (Sec.~\ref{SecEcEmp}) we demonstrate that  the indicators
behave similarly, indeed.

A similar observation was made in Ref.
\cite{ChakrabortiMarketCorr2003}. There the ``asset tree" was
studied, that is the recently introduced minimum spanning tree
description of correlations between stocks. The mean of the
normalized dynamic asset tree lengths was considered as a
promising indicator of the market dynamics. It appeared that a
simple average correlation coefficient gives the same signal in
time, as a more sophisticated indicator, the mean of the
normalized dynamic asset tree lengths. (compare Fig. 1 and Fig. 2
from Ref. \cite{ChakrabortiMarketCorr2003}). In Fig. 12 from that
paper very similar behavior of the mean correlation coefficient,
the normalized tree length, and the risk of the minimum risk
portfolio, as functions of time, was demonstrated.

In many publications in econophysics the average correlation
coefficient is used instead of the sums of absolute values in Eq.
(\ref{lpweight}). This is possible because in many financial
applications the orientation of the scales is fixed and the
difference between positive and negative correlations is very
important, for example, for portfolio optimization. In a more
general situation we have to use absolute values because we cannot
coordinate a priori the direction of different axes.

\section{Correlation and Risk in Physiology}

Effect of the simultaneous increase of the correlation and variance
under stress is supported by series of data from human physiology
and medicine.  In this Sec., we describe in brief several typical
examples. This is a review of already published experimental work.
More details are presented in an extended e-print
\cite{GorSmiTyuArX} and in original works.

\subsection{Data from Human Physiology}

\begin{figure} \centering{
a)\includegraphics[width=80mm]{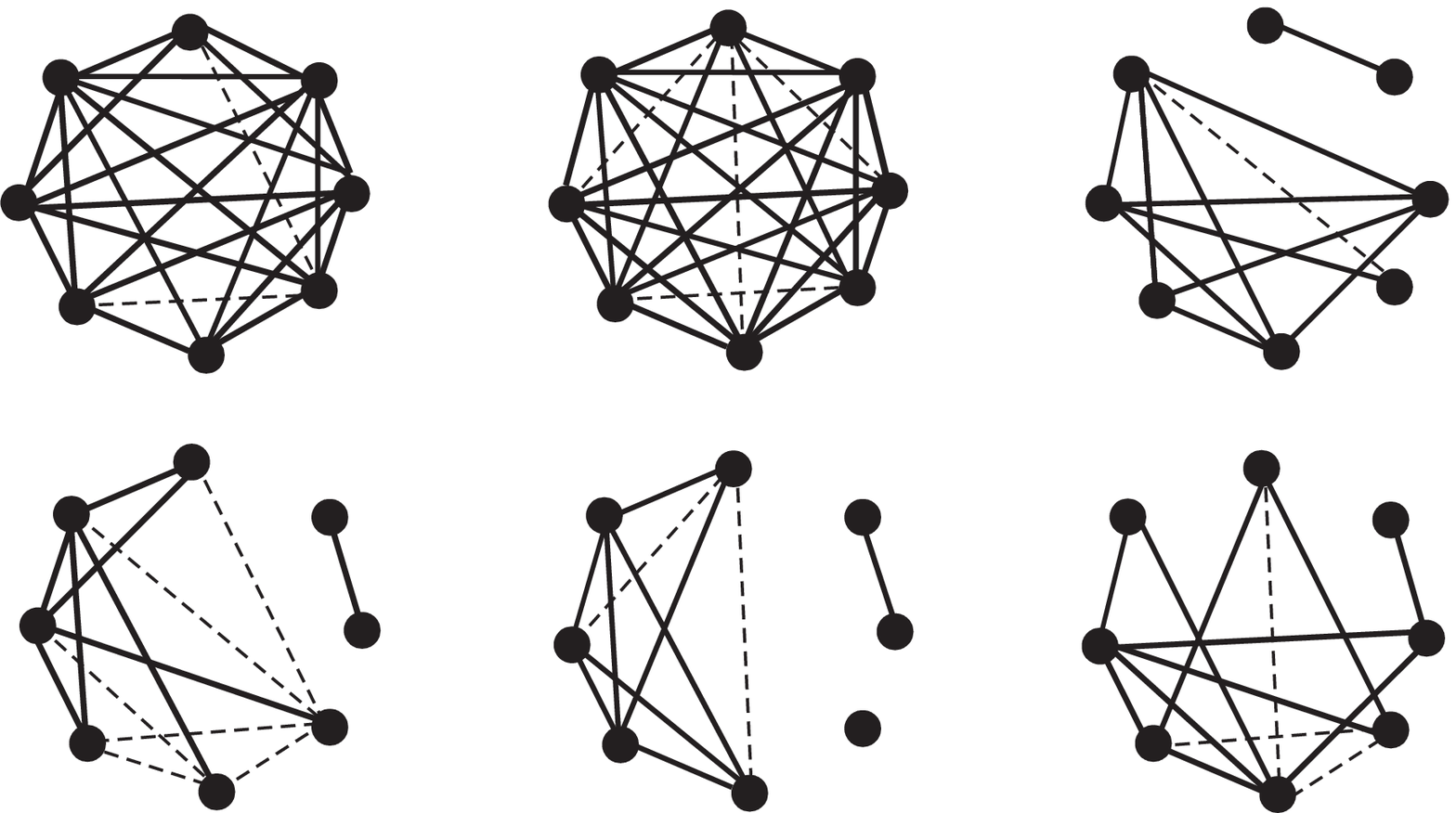}}
\hspace{10mm}b)\includegraphics[width=60mm]{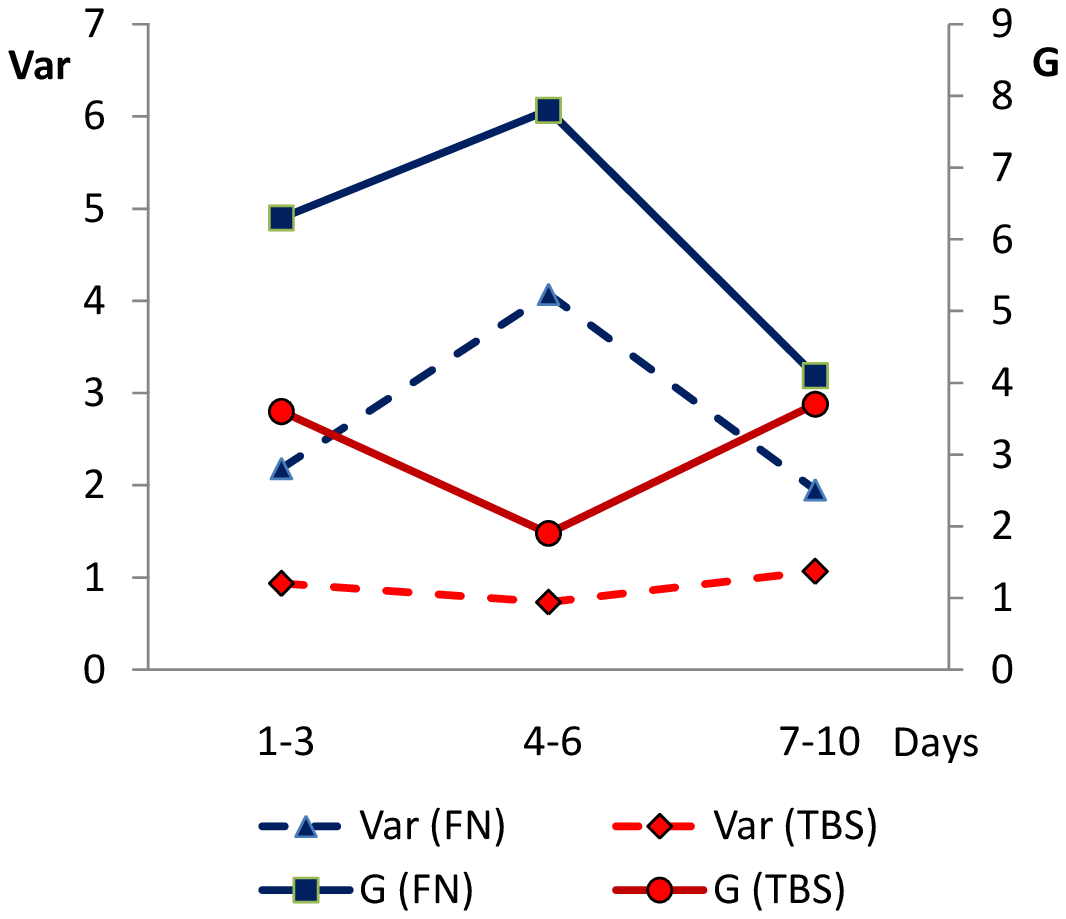}
\caption{\label{Fig:lipid1987} a) Correlation graphs of lipid
metabolism for newborn babies. Vertices correspond to different
fractions of lipids, solid lines correspond to correlation
coefficient between fractions $|r_{ij}| \geq 0.5$, dashed lines
correspond to $0.5 > |r_{ij}| \geq 0.25$. Upper row -- Far North
(FN), lower row -- the temperate belt of Siberia (TBS). From the
left to the right: 1st-3rd days (TBS -- 123  and FN -- 100
babies), 4th-6th days (TBS -- 98 and FN -- 99 babies), 7th-10th
days (TBS -- 35 and FN -- 29 babies). b) The weight of the
correlation graphs (solid lines) and the variance (dashed lines)
for these groups.}
\end{figure}

The first physiological system we studied in 1980s was the lipid
metabolism of healthy newborn babies born in the temperate belt of
Siberia (the comfort zone) and in the migrant families of the same
ethnic origin in a Far North city\footnote{The parents lived there
in standard city conditions.}. The blood analysis was taken in the
morning, on an empty stomach, at the same time each day. All the
data were collected during the summer. Eight lipid fractions were
analyzed \cite{GorSmiCorAd1st}. The resulting correlation graphs
are presented in Fig.~\ref{Fig:lipid1987}~a. Here solid lines
represent the correlation coefficient $|r_{ij}| \geq 0.5$, dashed
lines represent correlation coefficient $0.5 > |r_{ij}| \geq
0.25$. Variance monotonically increases with the weight of the
correlation graph (Fig.~\ref{Fig:lipid1987}~b).

Many other systems were studied. We analyzed the activity of
enzymes in human leukocytes during the short-term adaptation (20
days) of groups of healthy 20-30 year old men who change their
climate zone \cite{Bul1Limf,Bul2Limf}:
\begin{itemize}
\item{From the temperate belt of Siberia (Krasnoyarsk, comfort zone) to Far North in summer and in winter;}
\item{From Far North to the South resort (Sochi, Black Sea) in summer;}
\item{From the temperate belt of Russia to the South resort (Sochi, Black Sea) in summer.}
\end{itemize}
This analysis supports the basic hypothesis and, on the other
hand, could be used for prediction of the most dangerous periods
in adaptation, which need special care.

We selected the group of 54 people who moved to Far North, that
had any illness during the period of short-term adaptation.
After 6 months at Far North, this test group demonstrates much
higher correlations between activity of enzymes than the
control group (98 people without illness during the adaptation
period). We analyzed the activity of enzymes (alkaline
phosphatase, acid phosphatase, succinate dehydrogenase,
glyceraldehyde-3-phosphate dehydrogenase, glycerol-3-phosphate
dehydrogenase, and glucose-6-phosphate dehydrogenase) in
leucocytes: $G=5.81$ in the test group versus $G=1.36$ in the
control group. To compare the dimensionless variance for these
groups, we normalize the activity of enzymes to unite sample
means (it is senseless to use the trace of the covariance
matrix without normalization because normal activities of
enzymes differ in order of magnitude). For the test group. the
sum of the enzyme variances is 1.204, and for the control group
it is 0.388.

Obesity is a serious problem of contemporary medicine in developed
countries. The study was conducted on patients (more than 70
people) with different levels of obesity
\cite{RazzhevaikinObese2007}. The patients were divided into three
groups by the level of disease. Database with 50 attributes was
studied (blood morphology, cholesterol level including fractions,
creatinine, urea).

During 30 days patients received a standard treatment consisting
of a diet, physical activity, pharmacological treatment, physical
therapy and hydrotherapy. It was shown that the weight of the
correlation graph $G$ of more informative parameters was
originally high and monotonically dependent on the level of
sickness. It decreased during therapy.

\subsection{Data from Ecological Physiology of Plants}

The effect (Fig.~\ref{Fig:effect}) exists for plants too. It was
demonstrated, for example, by analysis of the impact of emissions
from a heat power station on Scots pine \cite{KofmantREES}. For
diagnostic purposes the secondary metabolites of phenolic nature
were used. They are much more stable than the primary products and
hold the information about the past impact of environment on the
plant organism for a longer time.

The test group consisted of 10 Scots pines (Pinus sylvestric L) in
a 40 year old stand of the II class in the emission tongue 10 km
from the power station. The station had been operating on brown
coal for 45 years. The control group of 10 Scots pines was from a
stand of the same age and forest type, growing outside the
industrial emission area. The needles for analysis were one year
old from the shoots in the middle part of the crown. The samples
were taken in spring in bud swelling period. The individual
composition of the alcohol extract of needles was studied by high
efficiency liquid chromatography.  26 individual phenolic
compounds were identified for all samples and used in the
analysis.

No reliable difference was found in the test group and control
group average compositions. For example, the results for
Proanthocyanidin content (mg/g dry weight) were as follows:
\begin{itemize}
\item{Total 37.4$\pm$3.2 (test) versus 36.8$\pm$2.0 (control);}
\end{itemize}
Nevertheless, the sample variance in the test group was 2.56 times
higher, and the difference in the correlations was huge: $G=17.29$
for the test group versus $G=3.79$ in the control group.

The grassy plants under trampling load also demonstrate a similar
effect \cite{RazzhevaikinTrava1996}. The grassy plants in an oak
forests are studied. For analysis, fragments of forests were
selected, where the densities of trees and bushes were the same.
The difference between these fragments was in damage to the soil
surface by trampling. The studied physiological attributes were:
the height of sprouts, the length of roots, the diameter of roots,
the amount of roots, the area of leaves, the area of roots. Again,
the weight of the correlation graph and the variance monotonically
increased with the trampling load.

\subsection{The Problem of ``No Return" Points}

It is practically important to understand where the system is
going: (i) to the bottom of the crisis with the possibility to
recover after that bottom, (ii) to the normal state, from the
bottom, or (iii) to the ``no return" point, after which it cannot
recover.

Situations between the comfort zone and the crisis has been
studied for dozens`of systems, and the main effect is supported
by much empirical evidence. The situation near the ``no return"
point is much less studied. Nevertheless, some observations
support the hypothesis presented for this case in
Fig.~\ref{Fig:effect}: when approaching the fatal situation
correlations decrease and variance increases.

This problem was studied with the analysis of fatal outcomes in
oncological \cite{MansurOnco} and cardiological \cite{Strygina}
clinics, and also in special experiments with acute hemolytic
anemia caused by phenylhydrazine in mice \cite{mice}. The main
result is: when approaching the no-return point, correlations
disappear ($G$ decreases), and variance typically does continue to
increase.

For example, the dynamics of correlations between physiological
parameters after myocardial infarction was studied in
Ref.~\cite{Strygina}. For each patient (more than 100 people),
three groups of parameters were measured: echocardiography-derived
variables (end-systolic and end-diastolic indexes, stroke index,
and ejection fraction), parameters of central hemodynamics
(systolic and diastolic arterial pressure, stroke volume, heart
rate, the minute circulation volume, and specific peripheral
resistance), biochemical parameters (lactate dehydrogenase, the
heart isoenzyme of lactate dehydrogenase LDH1, aspartate
transaminase, and alanine transaminase), and also leucocytes. Two
groups were analyzed after 10 days of monitoring: the patients
with a lethal outcome, and the patients with a survival outcome
(with compatible amounts of group members). These groups do not
differ significantly in the average values of parameters and are
not separable in the space measured attributes. Nevertheless, the
dynamics of the correlations in the groups are essentially
different. For the fatal outcome correlations were stably low
(with a short pulse at the 7th day), for the survival outcome, the
correlations were higher and monotonically grew. This growth can
be interpreted as return to the ``normal crisis" (the central
position in Fig.~\ref{Fig:effect}).

\begin{figure} \centering{
\includegraphics[width=80mm]{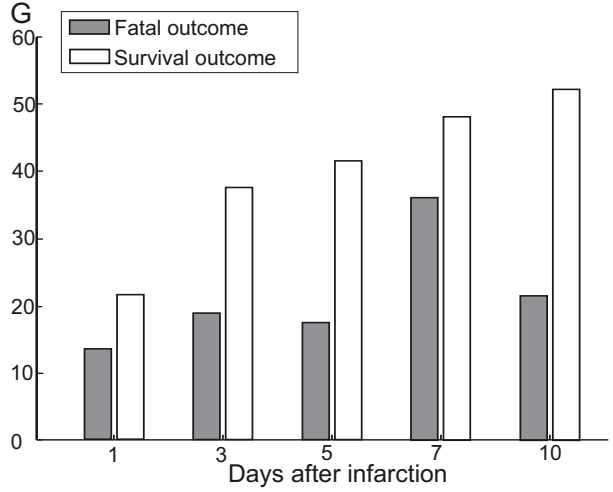}}
\caption{\label{Fig:Infarct}Dynamics of weight of the correlation
graphs of echocardiography-derived variables, parameters of
central hemodynamics, biochemical parameters, and also leucocytes
during 10 days after myocardial infarction for two groups of
patients: for the survival outcome and for the fatal outcome. Here
$G$ is the sum of the strongest correlations $|r_{ij}|>0.4$,
$i\neq j$ \cite{Strygina}.}
\end{figure}

Topologically, the correlation graph for the survival outcome
included two persistent triangles with strong correlations: the
central hemodynamics triangle, minute circulation volume -- stroke
volume -- specific peripheral resistance, and the heart
hemodynamics triangle, specific peripheral resistance -- stroke
index -- end-diastolic indexes. The group with a fatal outcome had
no such persistent triangles in the correlation graph.

In the analysis of fatal outcomes for oncological patients and in
special experiments with acute hemolytic anemia caused by
phenylhydrazine in mice one more effect was observed: for a short
time before death the correlations increased, and then fell down
(see also the pulse in Fig.~\ref{Fig:Infarct}). This short pulse
of the correlations (in our observations, usually for one day, a
day which precedes the fatal outcome) is opposite to the major
trend of the systems in their approach to death. We cannot claim
universality of this effect and it requires additional empirical
study.

\section{Correlations and Risk in Economics. Empirical Data \label{SecEcEmp}}

\subsection{Thirty Companies from the FTSE 100 Index. A Case Study}

\subsubsection{Data and Indicators}

For the analysis of correlations in financial systems  we used the
daily closing values over the time period 03.01.2006 -- 20.11.2008
for companies that are registered in the FTSE 100 index (Financial
Times Stock Exchange Index). The FTSE 100 is a
market-capitalization weighted index representing the performance
of the 100 largest UK-domiciled blue chip companies which pass
screening for size and liquidity. The index represents
approximately 88.03\% of the UK's market capitalization.  FTSE 100
constituents are all traded on the London Stock Exchange's SETS
trading system. We selected 30 companies that had the highest
value of the capital (on the 1st of January 2007) and  stand for
different types of business as well. The list of the companies and
business types is displayed in Table~\ref{CompList}.

\begin{table}\caption{Thirty largest companies for analysis
from the FTSE 100 index \label{CompList}} \centering{\small
\begin{tabular} {|c|l|l|l|}
  \hline
   Number &Business type & Company &  Abbreviation \\
  \hline \hline
  1& Mining & Anglo American plc& AAL\\
  2& & BHP Billiton& BHP\\
   \hline
  3& Energy (oil/gas) & BG Group & BG \\
  4& & BP & BP\\5&  & Royal Dutch Shell & RDSB \\
  \hline
  6 & Energy (distribution) & Centrica& CNA\\7& & National Grid & NG\\
  \hline
8&  Finance (bank) & Barclays plc & BARC\\9&   & HBOS & HBOS\\10& & HSBC HLDG & HSBC \\
 11& & Lloyds & LLOY\\
   \hline
  12& Finance (insurance)  & Admiral& ADM \\13& & Aviva & AV\\14& & LandSecurities&LAND\\
15&   & Prudential& PRU\\16& & Standard Chartered& STAN\\
    \hline
    17& Food production & Unilever &ULVR\\
    \hline
18& Consumer  & Diageo & DGE\\
19& goods/food/drinks & SABMiller& SAB\\
20& & TESCO &TSCO\\
\hline
21& Tobacco & British American Tobacco &BATS\\
22& & Imperial Tobacco & IMT\\
\hline
23& Pharmaceuticals& AstraZeneca &AZN\\
24& (inc. research)& GlaxoSmithKline & GSK\\
\hline
 25& Telecommunications & BT Group &BTA\\
 26& & Vodafone &VOD\\
\hline
27&Travel/leasure& Compass Group & CPG\\
\hline
28&Media (broadcasting) & British Sky Broadcasting & BSY\\
\hline
29& Aerospace/ & BAE System & BA\\
30& defence & Rolls-Royce& RR\\
\hline \hline
\end{tabular}
}\end{table}

Data for these companies are available form the Yahoo!Finance
web-site. For data cleaning we use also information for the
selected period available at the London Stock Exchange
web-site. Let $x_i(t)$ denote the closing stock price for the
$i$th company at the moment $t$, where $i=\overline{1,30}$,
$t=\overline{1,732}$. We analyze the correlations of
logarithmic returns: $x^l _i(t)=\ln \frac{x_i (t)} {x_i(t-1)}$,
$t=\overline{2,732}$ in sliding time windows of length $p=20$,
this corresponds approximately to 4 weeks of 5 trading days,
$t=\overline{p+1,732}$. The correlation coefficients
$r_{ij}(t)$ and all indicators for time moment $t$ are
calculated in the time window $[t-p,t-1]$, which precedes $t$.
This is important if we would like to consider changes in these
indicators as precursors of crisis.

The information about the level of correlations could be represented
in several ways. Here we compare 4 indicators:
\begin{itemize}
  \item The non-diagonal part of the correlation matrix in $L_2$ norm -
  $\|r\|_2$;
  \item The non-diagonal part of the correlation matrix in $L_1$ norm -
  $\|r\|_1$;
  \item The sum of the strongest elements
  $G=\sum_{j>i,|r_{ij}|>0.5}|r_{ij}|$;
  \item The amount Dim of principal components estimated due to the broken stick
  model.
\end{itemize}

\begin{figure}\centering{
\includegraphics[width=130mm]{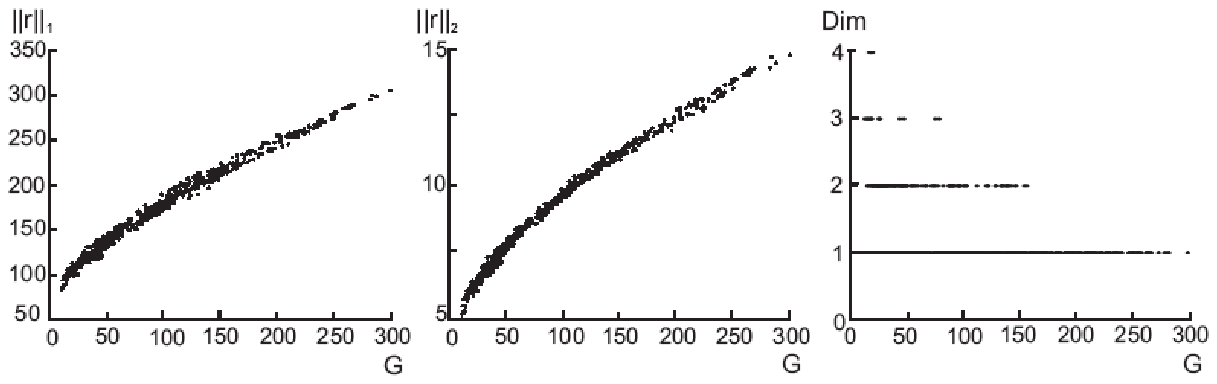}}
\caption{Scatter diagrams for three pairs of indicators:
$G-\|r\|_1$, $G-\|r\|_2$, and $G-$Dim, where Dim is amount of
principal components  estimated due to the broken stick model.
\label{scatterDiagrams}}
\end{figure}

The dynamics of the first three indicators are quite similar.
Scatter diagrams (Fig.~\ref{scatterDiagrams}) demonstrate a strong
connection between the indicators. We used the weight of the
correlation graph $G$ (the sum of the strongest correlations
$r_{ij}>0.5$, $i\neq j$) for our further analysis.

\begin{figure}\centering{
\includegraphics[width=100mm]{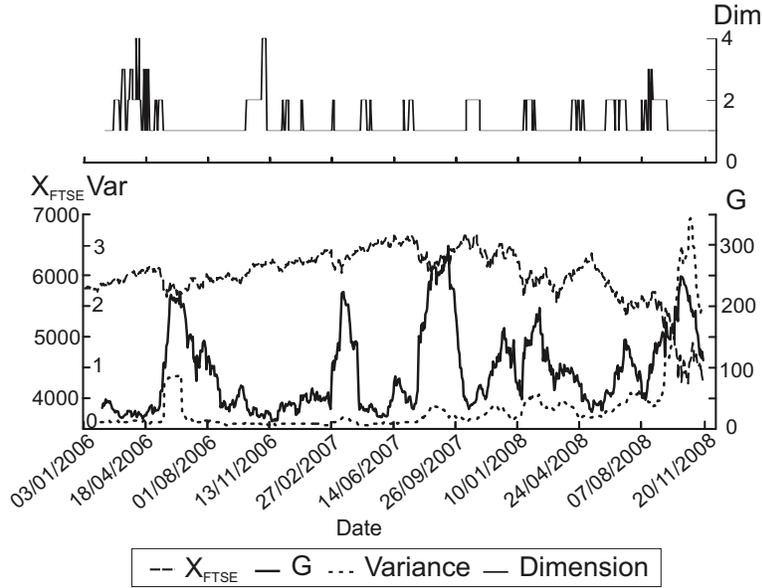}}
\caption{Dynamics of FTSE index, G, Variance, and Dimension
estimated  due to the broken stick model. \label{FTSE_G05_Dim}}
\end{figure}

Fig.~\ref{FTSE_G05_Dim}  allows us to compare dynamics of
correlation, dimension and variance  to the value of FTSE100.
Correlations increase when the market goes down and decrease when it
recovers. Dynamics of variance of log-returns has the same tendency.
To analyze the critical periods in more detail, let us select
several time intervals and several positions of the sliding window
inside these intervals.

\subsubsection{Correlation Graphs for Companies}

\begin{figure}\centering{
a)\includegraphics[width=80mm]{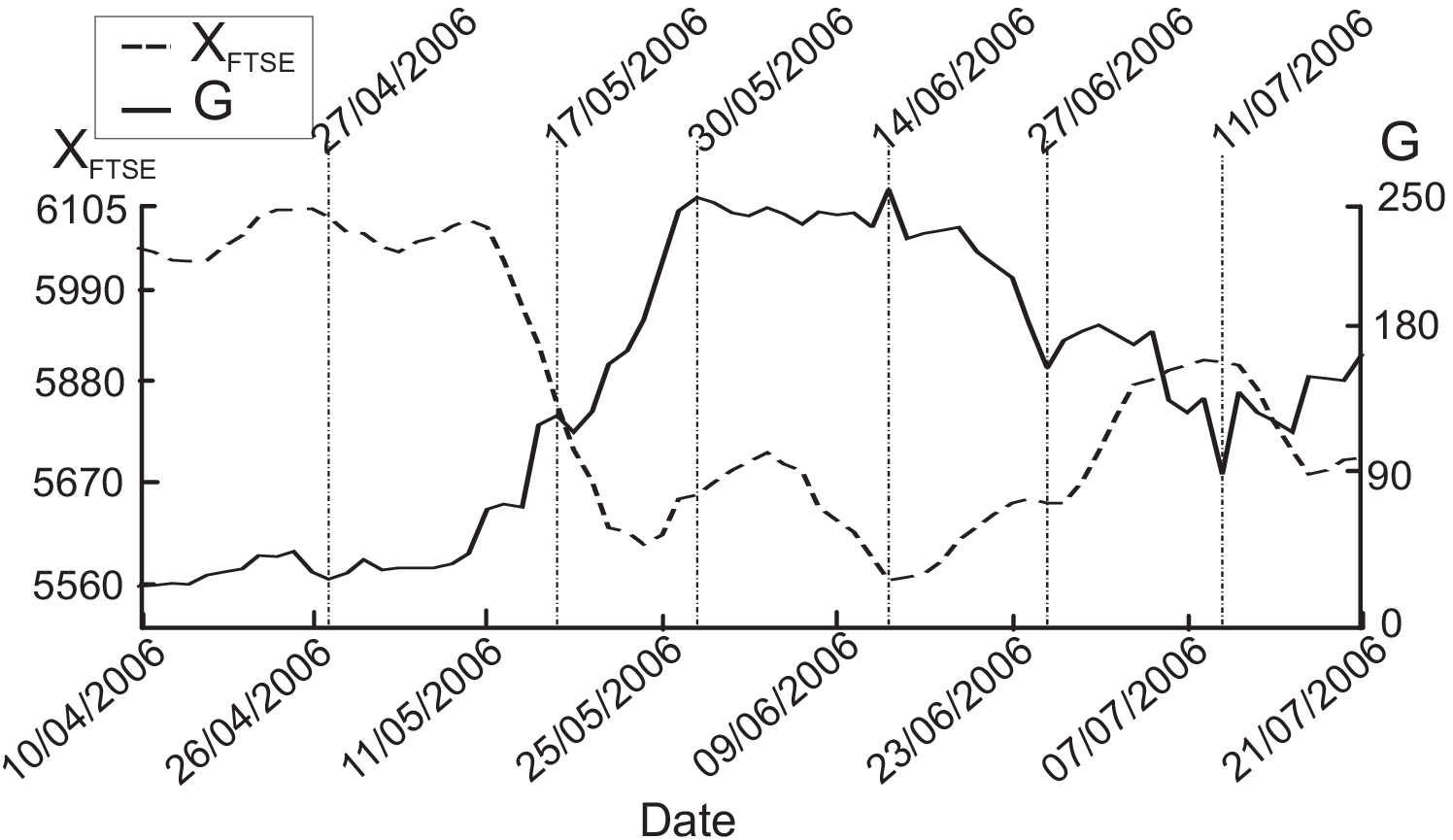}b)\includegraphics[width=50mm]{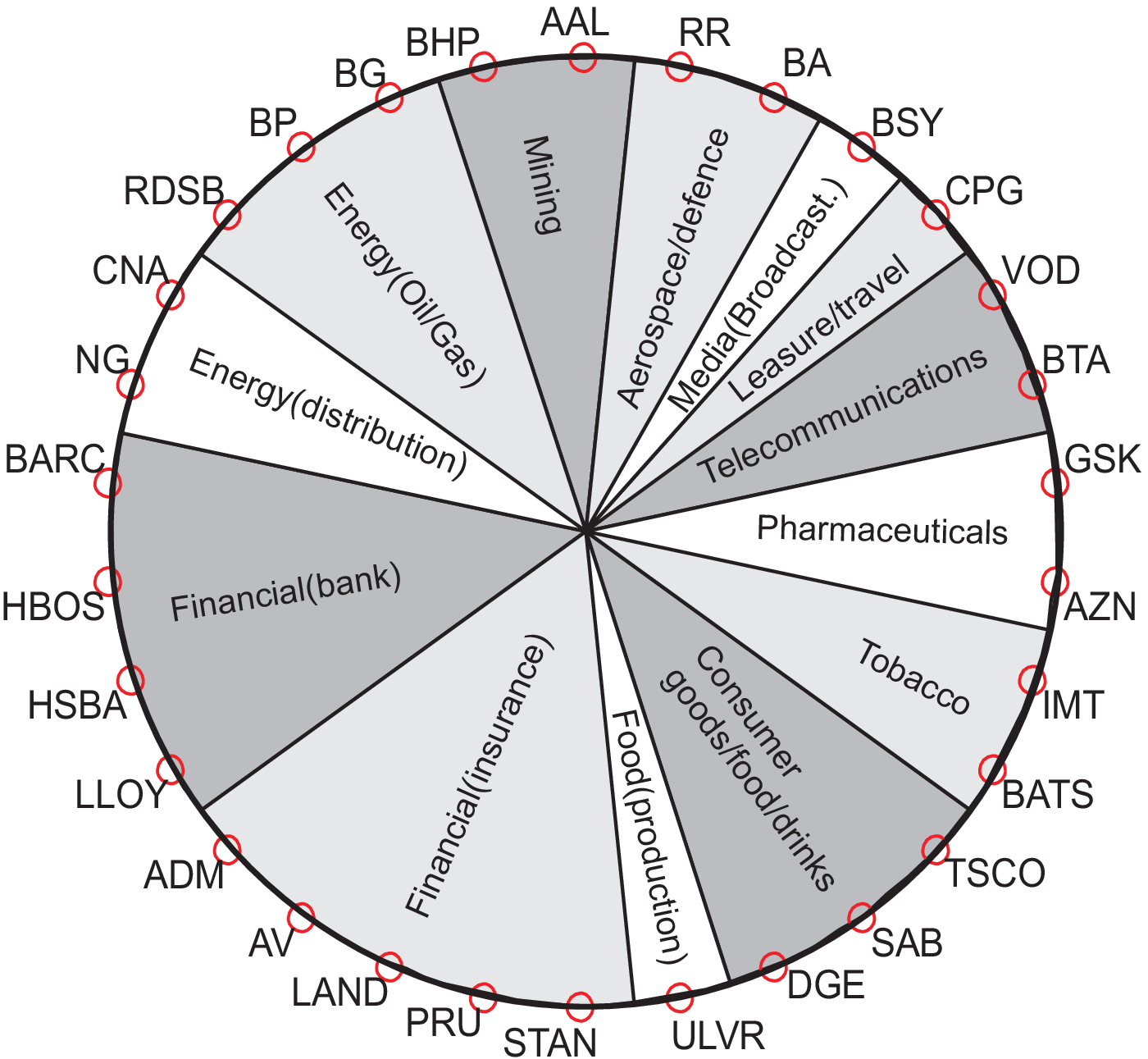}\\
 c)\includegraphics[width=130mm]{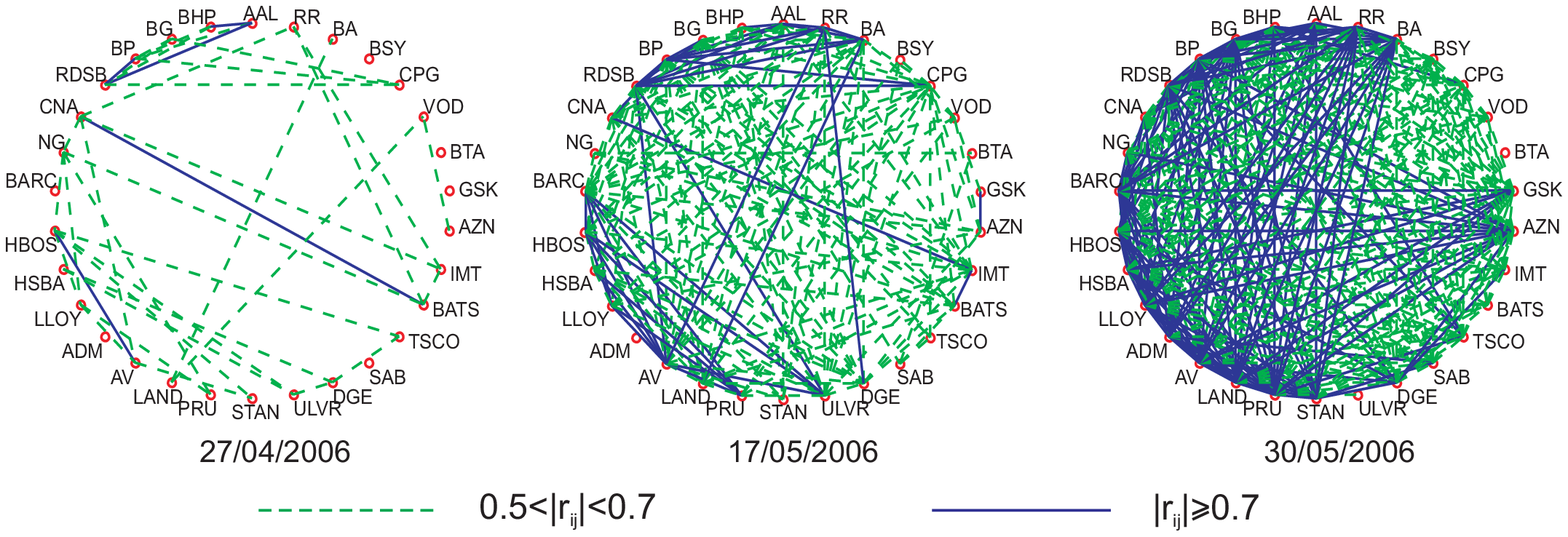} \\
 d)\includegraphics[width=130mm]{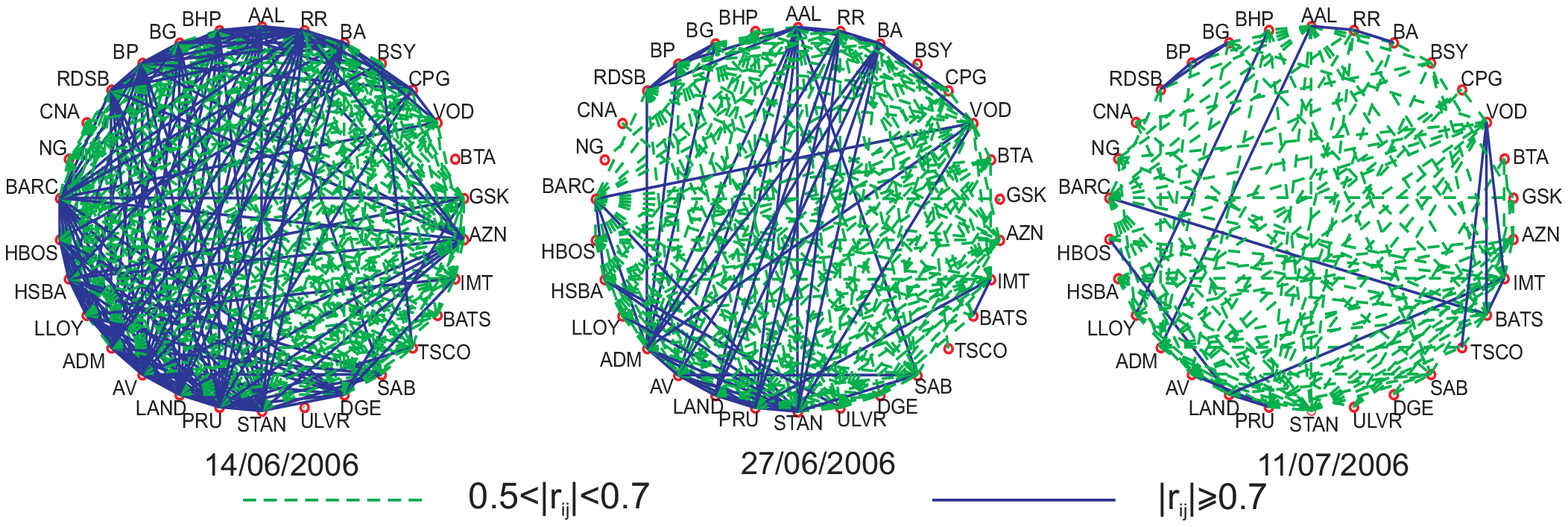}}
\caption{Correlation graphs for six positions of the sliding time
window
 on interval 10/04/2006 - 21/07/2006.
a)~Dynamics of FTSE100 (dashed line) and of $G$ (solid line) over
the interval, vertical lines correspond to the points that were used
for the correlation graphs.
 b)~Thirty companies for analysis and their distributions over
various sectors of economics.
 c)~The correlation graphs for the
first three points, FTSE100 decreases, the correlation graph becomes
more connective. d)~The correlation graphs for the last three
points, FTSE100 increases, the correlation graph becomes less
connective.\label{graph_int1}}
\end{figure}

\begin{figure}\centering{
a)\includegraphics[width=80mm]{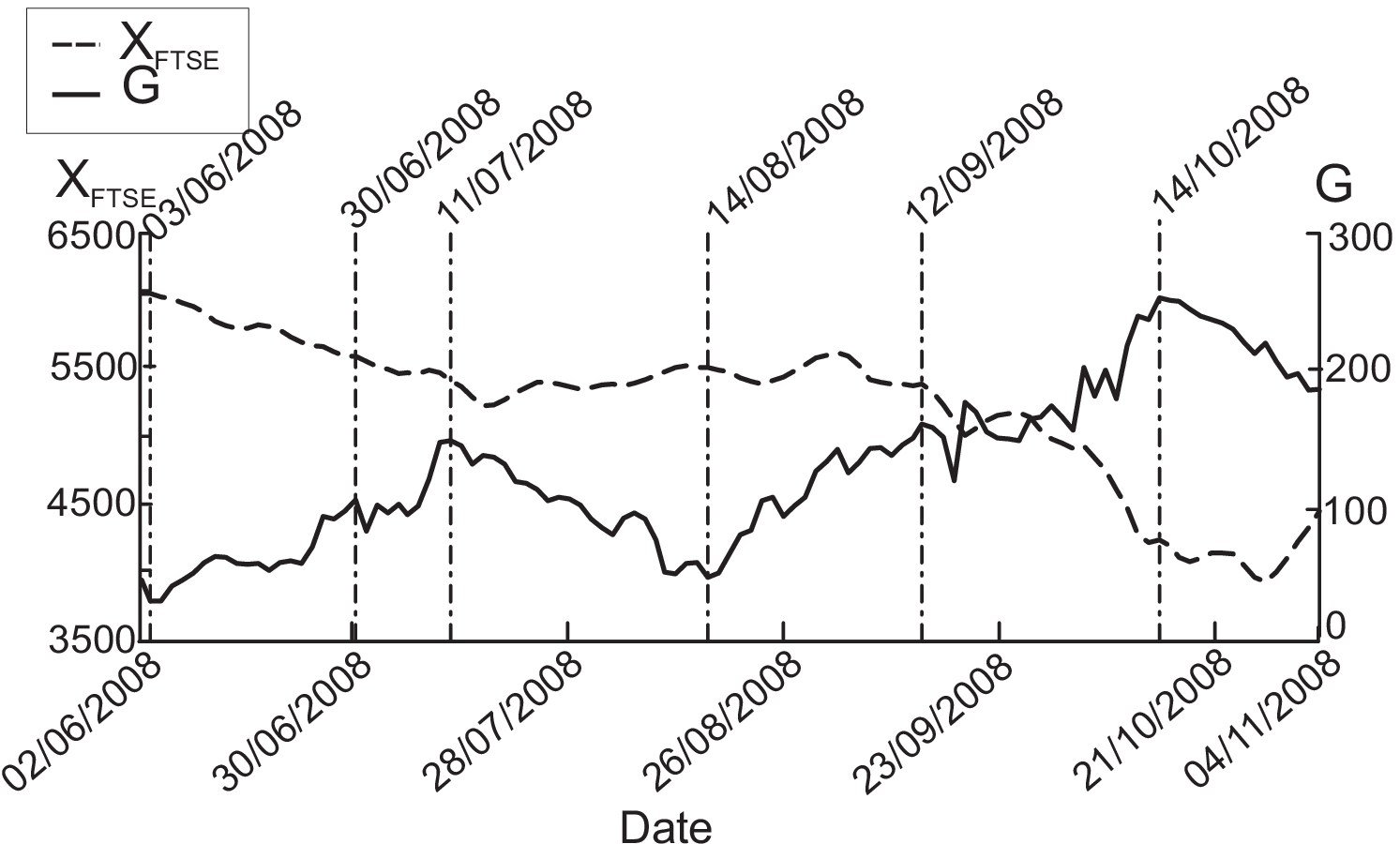}b)\includegraphics[width=50mm]{Pie_business.eps}\\
c)\includegraphics[width=130mm]{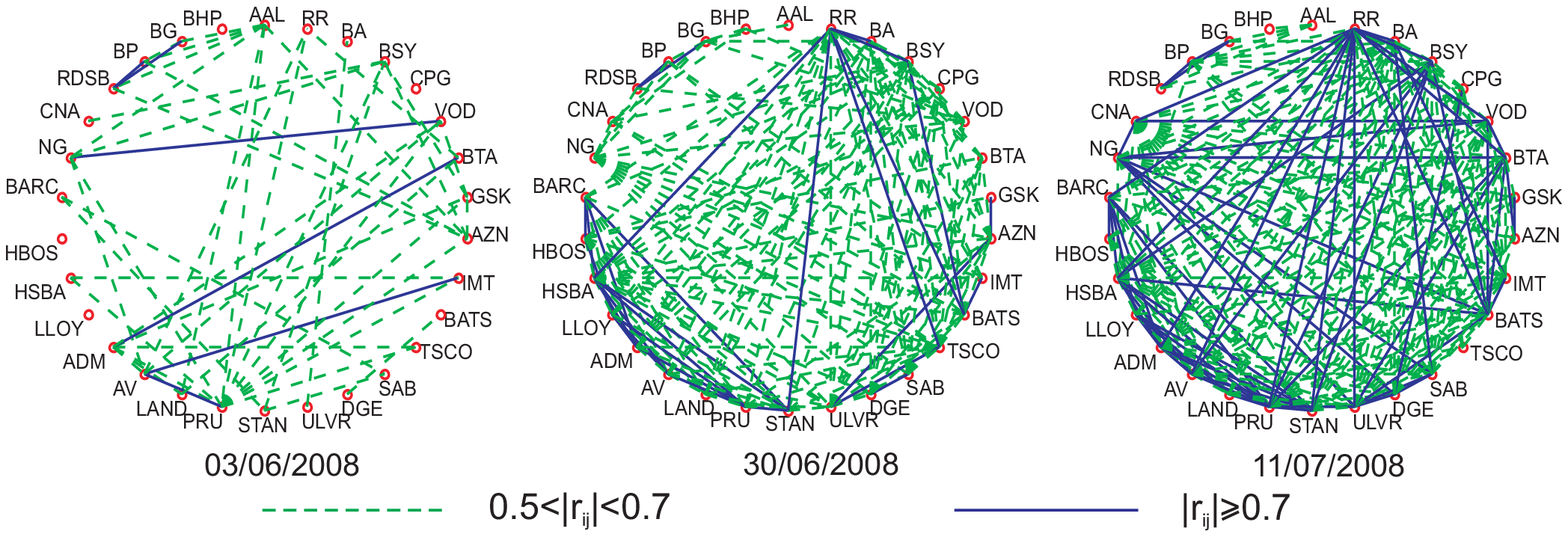}\\
d)\includegraphics[width=130mm]{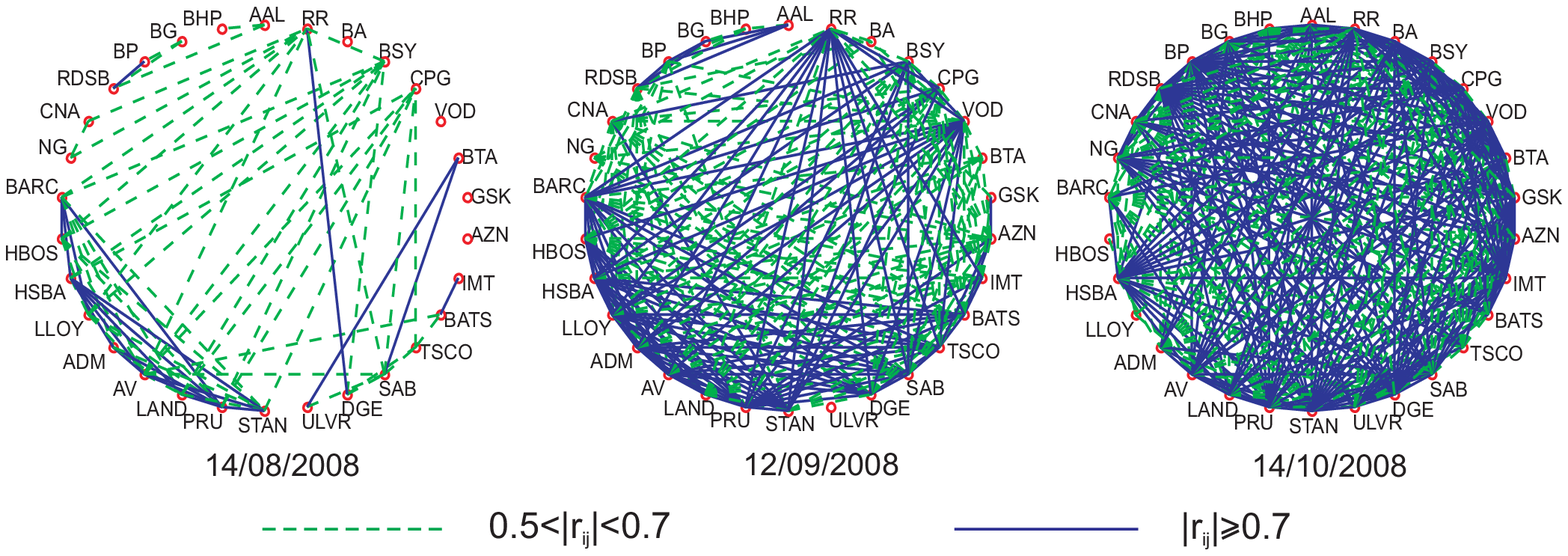}}
\caption{\label{graph_int3} Correlation graphs for six positions
of the sliding time window
 on interval 02/06/2008 - 04/11/2008.
 a)~Dynamics of FTSE100 (dashed line) and of $G$ (solid line) over the interval, vertical lines
correspond to the points that were used  for the correlation graphs.
 b)~Thirty companies for analysis and their distributions over
various sectors of economics.
 c)~The correlation graphs for the first three points, FTSE100
decreases, the correlation graph becomes more connective. Between
the third and the 4th points FTSE100 increased, and the first graph
here is more rarefied than at the third point. Between the third and
the 4th points FTSE100 slightly increased, correlation decreased,
and the first graph at the next row is more rarefied than at the
third point.
 d)~The correlation
graphs for the last three points, FTSE100 decreases, the correlation
graph becomes more connective.}
\end{figure}

We extracted 2 intervals for more detailed analysis. The first
interval, 10/04/2006 - 21/07/2006, represents the FTSE index
decrease and restoration in spring and summer 2006. The second
interval, 02/06/2008 - 04/11/2008, is a part of the financial
crisis. In each interval we selected six points and analyzed the
structure of correlations for each of these points (for a time
window, which precedes this point). For each selected point, we
create a correlation graph, solid lines represent correlation
coefficient $|r_{ij}| \geq \sqrt{0.5}$
($\sqrt{0.5}=\cos(\pi/4)\approx 0.707$), dashed lines represent
correlation coefficient $\sqrt{0.5} > |r_{ij}| \geq 0.5$:
(Figs.~\ref{graph_int1}c,d, \ref{graph_int3}c,d). On these
correlation graphs it is easy to observe, how critical correlations
appear, how are they distributed between different sectors of
economics, and how the crisis  moves from one sector to another.

There is no symmetry between the periods of the FTSE index decrease
and recovering. For example, in Fig.~\ref{graph_int1}c we see that
at the beginning (falling down) the correlations inside the
financial sector are important and some correlations inside industry
are also high, but in the corresponding recovering period
(Fig.~\ref{graph_int1}d) the correlations between industry and
financial institutions become more important.

All the indicators demonstrate the most interesting behavior at
the end of 2008 (Fig.~\ref{FTSE_G05_Dim}). The growth of variance
in the last peak is extremely high, but the increase of
correlations is rather modest. If we follow the logic of the basic
hypothesis (Fig.~\ref{Fig:effect}), then we should suspect that
the system is going to ``the other side of crisis", not to
recovery, but to disadaptation, this may be the most dangerous
symptom.

\subsubsection{Graphs for Correlations in Time}

The vector of attributes that represents a company is a 20 day
fragment of the time series. In standard biophysical research, we
studied correlations between attributes of an individual, and
rarely, correlation between individuals for different attributes. In
econophysics the standard situation is opposite. Correlation in time
is evidenced to be less than correlation between companies
\cite{VarVolLillo2000}. Nevertheless, correlation between days in a
given time window may be a good indicator of crisis.

\begin{figure}\centering{
\includegraphics[width=100mm]{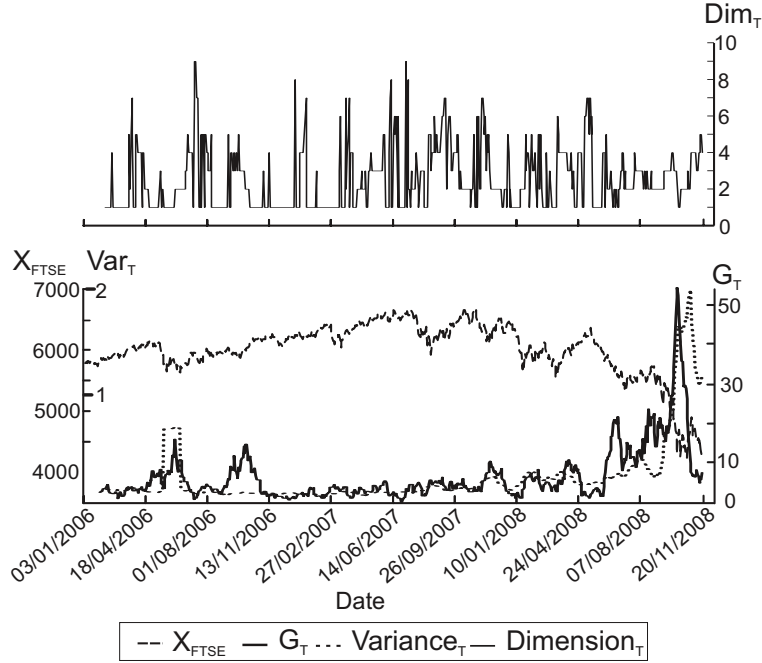}}
\caption{Dynamics of the market $X_{FTSE}$, weight of correlation
$G_T$ the sum of the correlation coefficients with absolute value
greater then $0.25$, Variance (volatility), and dimension of the
correlation matrix  estimated due to the broken stick model.
\label{FTSE_Gt_Dim}}
\end{figure}

Let us use here $G_T$ for the weight of the correlation graph in
time. Because correlation in time is less than between stocks, we
select here another threshold: $G_T$ is the sum of the correlation
coefficients with absolute value greater then $0.25$. FTSE
dynamics together with values of $G_T$ are presented in
Fig.~\ref{FTSE_Gt_Dim}. Solid lines represent a correlation
coefficient $|r_{ij}| \geq 0.5$, dashed lines represent a
correlation coefficient $0.5 > |r_{ij}| \geq 0.25$.

\begin{figure}\centering{
a)\includegraphics[width=90mm]{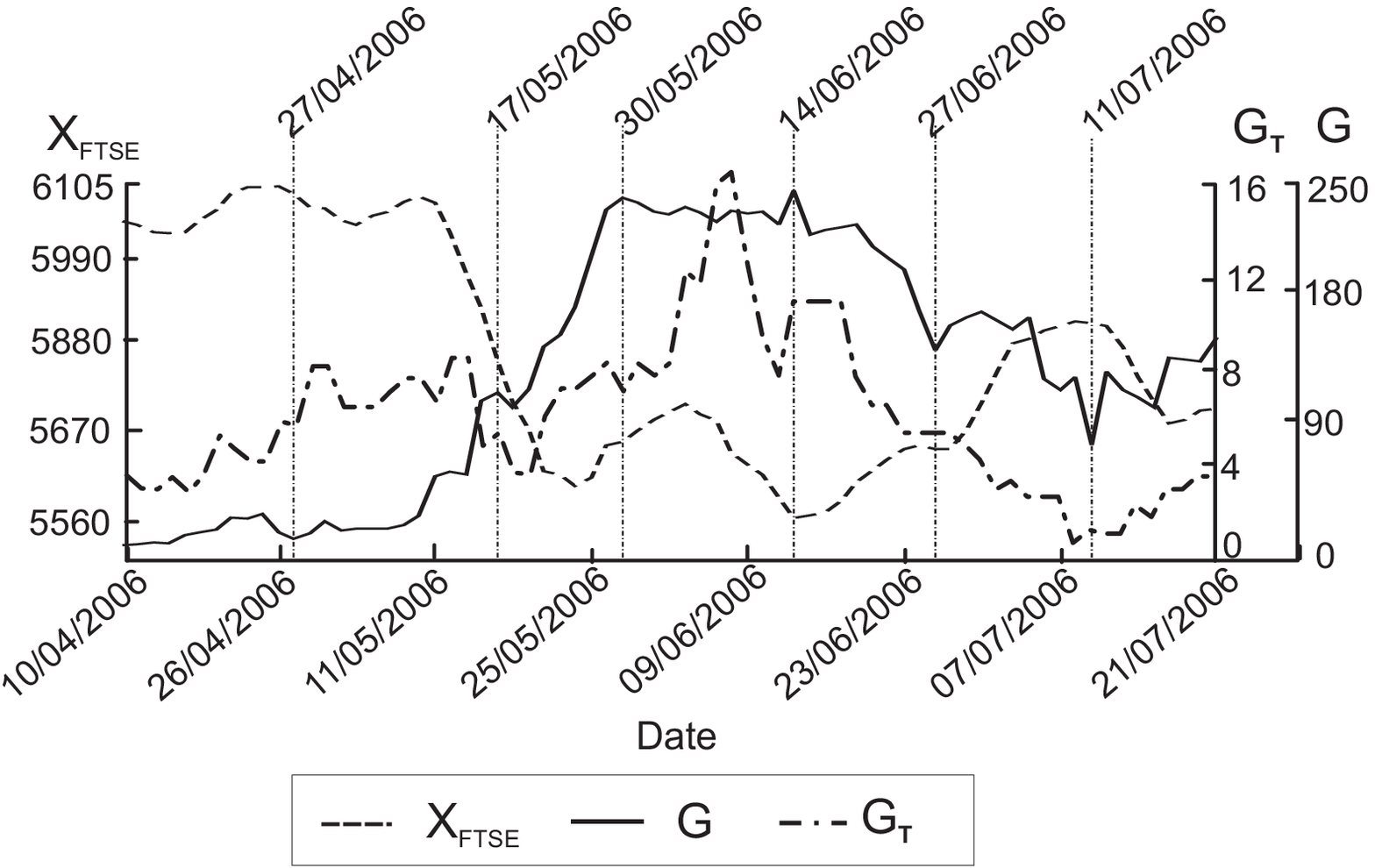}\\
 b)\includegraphics[width=130mm]{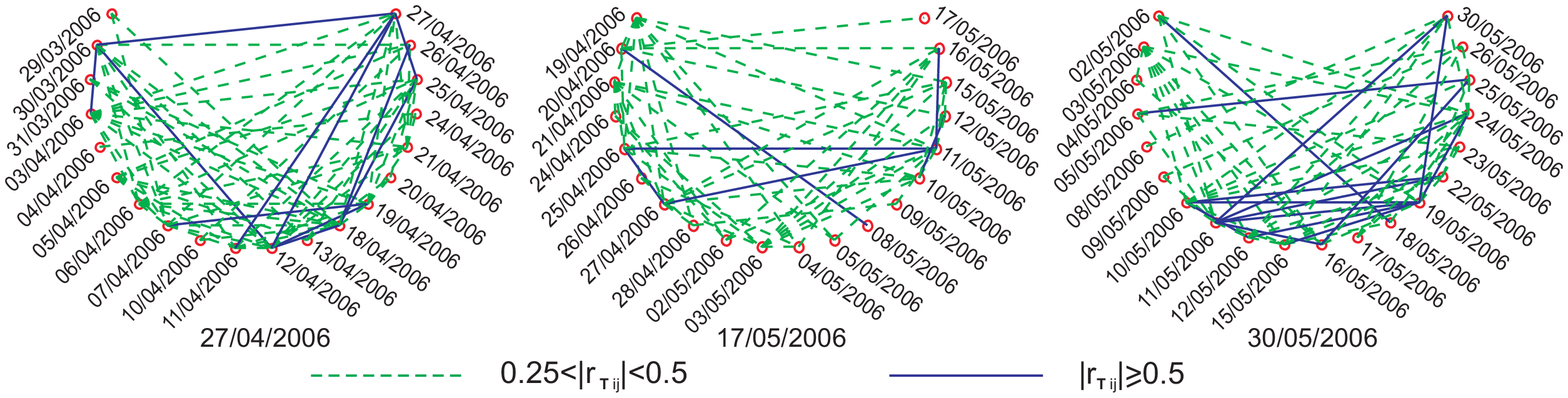} \\
 c)\includegraphics[width=130mm]{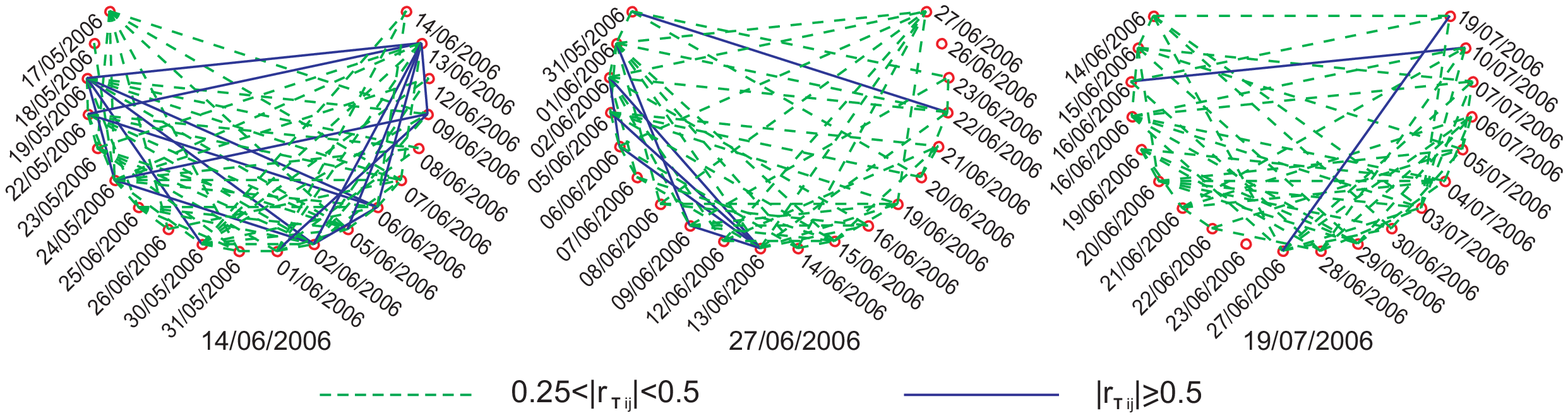}}
\caption{Graphs for correlation in time for six positions of the
sliding time window on interval 10/04/2006 - 21/07/2006.
 a)~Dynamics of
FTSE100 (dashed line), $G$ (solid line)  and $G_T$ (dash-and-dot
line) over the interval, vertical lines correspond to the points
that were used for the correlation graphs.
 b)~The correlation graphs for the
first three points: FTSE100 decreases and the correlation graph
becomes more connective.
 c)~The correlation graphs for the last three
points: FTSE100 increases and the correlation graph becomes less
connective.\label{Time_graph_int1}}
\end{figure}

\begin{figure}\centering{
a)\includegraphics[width=90mm]{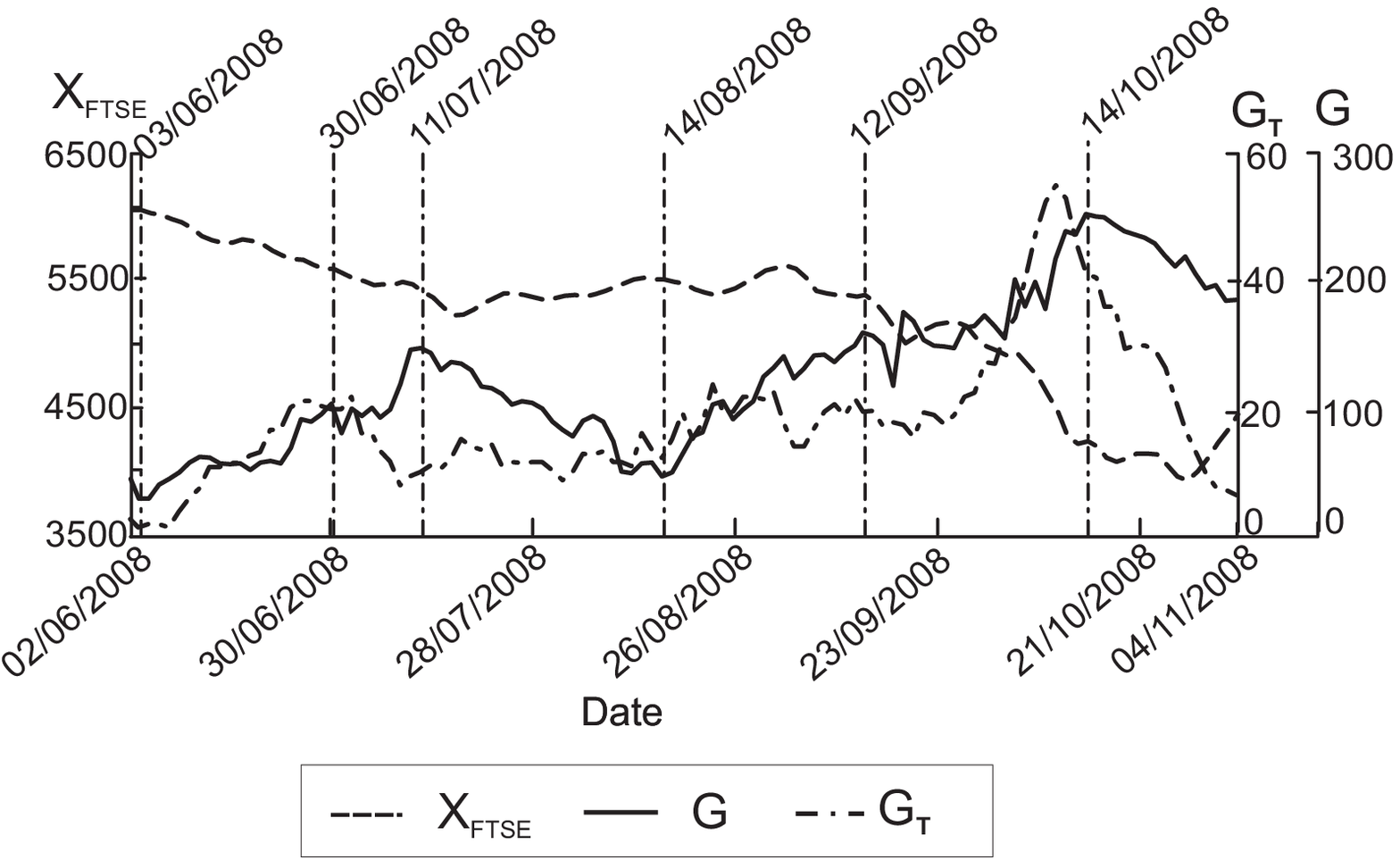}\\
b)\includegraphics[width=130mm]{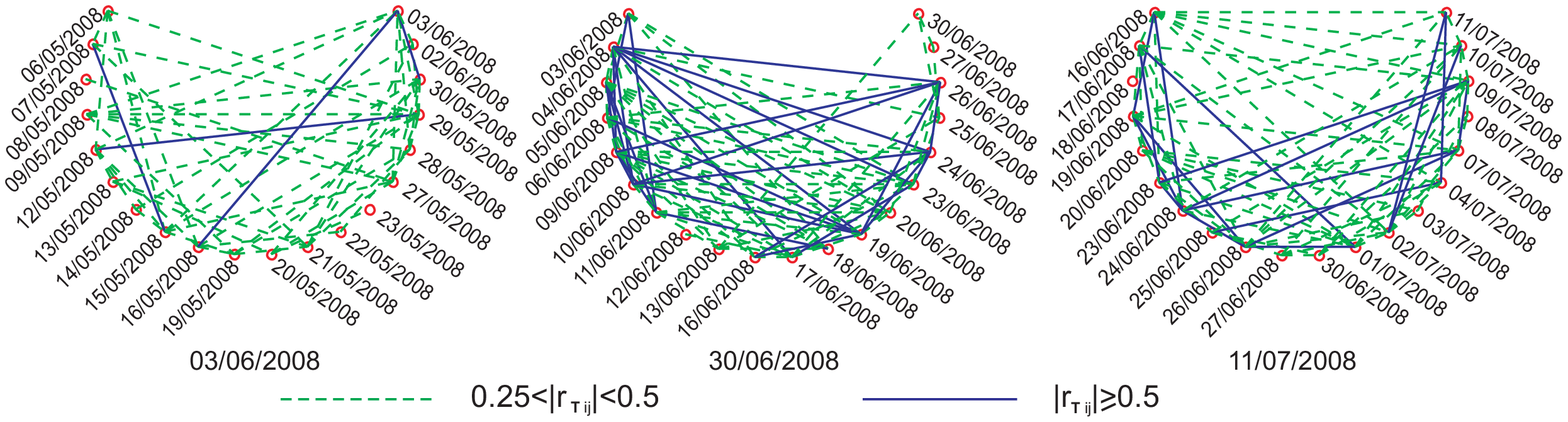}\\
c)\includegraphics[width=130mm]{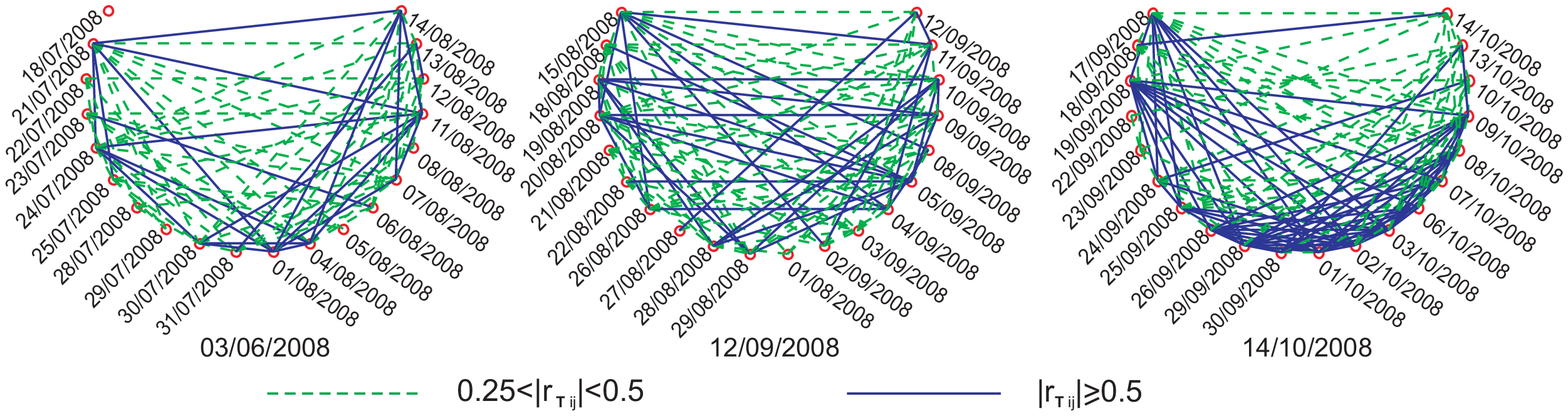}}
\caption{\label{Time_graph_int3} Correlation graphs for six
positions of the sliding time window
 on interval 02/06/2008 - 04/11/2008.
 a)~Dynamics of FTSE100 (dashed line), $G$ (solid line)  and $G_T$ (dash-and-dot
line) over the interval, vertical lines correspond to the points
that were used  for the correlation graphs.
 b)~Thirty companies for analysis and their distributions over
various sectors of economics.
 c)~The correlation graphs for the first three points: FTSE100
decreases and the correlation graph becomes more connective. Between
the third and the 4th points FTSE100 increases, and the first graph
here is more rarefied than at the third point. Between the third and
the 4th points FTSE100 slightly increases, correlation decreased,
and the first graph at the next row is more rarefied than at the
third point.
 d)~The correlation
graphs for the last three points: FTSE100 decreases and the
correlation graph becomes more connective.}
\end{figure}

On the figures \ref{Time_graph_int1}, \ref{Time_graph_int3} we
combined graphs of days correlations - 20 trading days prior to the
selected days.

An analysis of the dynamics of $G_T$ allows us to formulate a
hypothesis: typically, after the increase of $G_T$ the decrease
of FTSE100 index follows (and, the decrease of $G_T$ precedes
the increase of FTSE100). The time delay is approximately two
working weeks. In that sense, the correlation in time seems to
be better indicator of the future change, than the correlation
between stocks which has no such time gap. On the other hand,
the amplitude of change of $G_T$ is much smaller, and some of
the decreases of the FTSE100 index could not be predicted by
increases of $G_T$ (Fig.~\ref{FTSE_Gt_Dim}).

These observations are still preliminary and need future analysis
for different financial time series.

A strong correlation between days appears also with some time gap:
the links emerge, not usually between nearest days, but mostly
with an interval 4-15 days (see Figs.~\ref{Time_graph_int1},
\ref{Time_graph_int3}).

\subsection{Correlations and Crisis in Financial Time Series}

In economics and finance, the correlation matrix is very important
for the practical problem of portfolio optimization and
minimization of risk. Hence, an important problem arises: are
correlations constant or not? The hypothesis about constant
correlations was tested for monthly excess returns for seven
countries (Germany, France, UK, Sweden, Japan, Canada, and US)
over the period 1960-90 \cite{LonginCorrNonconst1995}. Correlation
matrices were calculated over a sliding window of five years. The
inclusion of October 1987 in the window led to an increase of
correlation in that window. After an analysis of correlations in
six periods of five years the null hypothesis of a constant
correlation matrix was rejected. In addition, the {\it conditional
correlation matrix} was studied. The multivariate process for
asset return was presented as
\begin{equation}
R_t=m_{t-1}+e_t; \; \; m_{t-1}=\mathbf{E}(R_t|F_{t-1}),
\end{equation}
where $R_t$ is a vector of asset returns and $m_{t-1}$ is the
vector of expected returns at time $t$ conditioned on the
information set $F_{t-1}$ from the previous step. Vector $e_t$ is
the unexpected (unpredicted) component in the asset returns.
Correlations between its components are called {\it conditional
correlations}. It was demonstrated that these conditional
correlations are also not constant. Two types of change were
found. Firstly, the correlations have a statistically significant
time trend and grow in time. The average increase in correlation
over 30 years is 0.36. Secondly, correlations in periods of high
volatility (high variance) are higher. To obtain this result, the
following model for the correlation coefficient was identified:
\begin{equation}
r^{i,{\rm us}}_t=r^{i,{\rm us}}_0 +r^{i,{\rm us}}_1 S_t^{\rm us},
\end{equation}
where $r^{i,{\rm us}}_t$ is the correlation coefficient between the
unexpected (unpredicted) components in the asset returns for the
$i$th country and the US, $S_t$ is a dummy variable that takes the
value 1 if the estimated conditional variance of the US market for
time $t$ is greater than its unconditional (mean) value and 0
otherwise. The estimated coefficient $r_1$ is positive for all
countries. The average over all countries for $r_0$ is equal to
0.430, while the average {\it turbulence effect} $r_1$ is 0.117
\cite{LonginCorrNonconst1995}. Finally, it was demonstrated that
other informational variables can explain more changes in
correlations than just the ``high volatility -- low volatility"
binning.

To analyze correlations between European equity markets before and
after October 1987, three 76-month periods were compared: February
1975--May 1981, June 1981--September 1987, and November
1987--February 1994 \cite{MericEuMarket1987}. The average
correlation coefficient for 13 equity markets (Europe + US)
increased from 0.37 in June 1981--September 1987 to 0.5 in
November 1987--February 1994. The amount of significant principal
components selected  by Kaiser's rule decreases from 3 (in both
periods before October 1987) to 2 (in the period after October
1987) for all markets and even from 3 to 1 for 12 European markets
\cite{MericEuMarket1987}. Of course, in average values for such
long periods it is impossible to distinguish the consequences of
the October 1987 catastrophe and a trend of correlation
coefficients (that is, presumably, nonlinear).

Non-stationarity of the correlation matrix was demonstrated in a
detailed study of the financial empirical correlation matrix of
the 30 companies which Deutsche Aktienindex (DAX) comprised during
the period 1988--1999 \cite{DrozdComCollNoise2000}. The time
interval (time window) is set to 30 and continuously moved over
the whole period. It was evidenced that the drawups and the
drawdowns of the global index (DAX) are governed,  respectively,
by dynamics of a significantly distinct nature. The drawdowns are
dominated by one strongly collective eigenstate with a large
eigenvalue. The opposite applies to drawups: the largest
eigenvalue moves down which is compensated by a simultaneous
elevation of lower eigenvalues. Distribution of correlation
coefficients for these data have a distinctive bell-like shape
both for one time window (inside one correlation matrix) and for
ensemble of such sliding windows in a long time period.

This observation supports the idea of applying the theory of the
Gaussian matrix ensembles to the analysis of financial time
series. The random matrix theory gives a framework for analysis of
the cross-correlation matrix for multidimensional time series. In
that framework, stock price changes of the largest 1000 U.S.
companies were analyzed for the 2-year period 1994--1995
\cite{Stanley1999}, and statistics of several of the largest
eigenvalues was evidenced to be far from the random matrix
prediction, but the distribution of ``the rest" of the eigenvalues
and eigenvectors satisfies the random matrix ensemble. The crucial
question is: where is the border between the random and the
non-random parts of spectra? Formula (\ref{RMTeiganvalues}) gives
in this case $\lambda_{\max}\approx 2$. The random matrix theory
predicts for the Gaussian orthogonal ensembles that the components
of the normalized eigenvectors are distributed according to a
Gaussian probability distribution with mean zero and variance one.
Eigenvectors corresponding to most eigenvalues in the ``bulk"
($\lambda< 2$) have the Gaussian distribution, but eigenvectors
with bigger eigenvalues significantly deviate from this.
\cite{Stanley1999}.

This kind of analysis was continued for the three major US stock
exchanges, namely the New York Stock Exchange (NYSE), the American
Stock Exchange (AMEX), and the National Association of Securities
Dealers Automated Quotation (NASDAQ) \cite{Stanley2002}. The concept
of ``deviating eigenvectors" was developed, these vectors correspond
to the eigenvalues which are systematically outside the random
matrices ensembles predictions. Analysis of ``deviating
eigenvectors" which are outside the random matrices ensembles
predictions (\ref{RMTeiganvalues}) gives information of major
factors common to all stocks, or to large business sectors. The
largest eigenvalue was identified as the ``market mode". During
periods of high market volatility values of the largest eigenvalue
are large.  This fact was commented as a strong collective behavior
in regimes of high volatility. For the largest eigenvalue, the
distribution of coordinates of the eigenvector has very remarkable
properties:
\begin{itemize}
\item{It is much more uniform than the prediction of the random matrix
theory (authors of Ref.~\cite{Stanley2002} described this vector as
``approximately uniform", suggesting that all stocks participate in
this ``market mode");}
\item{Almost all components of that eigenvector have the same sign.}
\item{A large degree of cross correlations between stocks can
be attributed to the influence of the largest eigenvalue and its
corresponding eigenvector}
\end{itemize}
Two interpretations of this eigenvector were proposed: it
corresponds either to the common strong factor that affects all
stocks, or it represents the ``collective response" of the entire
market to stimuli.

Spectral properties of the correlation matrix were analyzed
also for 206 stocks traded in the Istanbul Stock Exchange
Market during the 5-year period 2000--2005
\cite{SadikCrosCorFin2007}. One of the main results of this
research is the observation that the correlations among stocks
are mostly positive and tend to increase during crises. The
number of significant eigenvalues (outside the random matrix
interval) is smaller than it was found in previous study of the
well-developed international market in the US. The possible
interpretation is: the emerging market is ruling by smaller
amount of factors.

An increase of correlations in a time of crisis was demonstrated
by the analysis of 150 years of market dynamics
\cite{Goetzmann2004}. As a result, in the year 2004 it was
mentioned very optimistically: ``Our tests suggest that the
structure of global correlations shifts considerably through time.
It is currently near an historical high -– approaching levels of
correlation last experienced during the Great Depression".
Nevertheless, it remains unclear, does the correlation cause the
transmission chain of collapse or is it inextricably tied to it
\cite{Smith2009}?

There are several types of explanation of these correlation
effects. One can look for the specific reasons in the balance
between specialization and globalization, in specific fiscal,
monetary, legal, cultural or even language conditions, in dynamics
of fundamental economic variables such as interest rates and
dividend yields, in the portfolio optimization by investors, and
in many similar more or less important processes. These specific
explanations could work, but for such a general effect it is
desirable to find a theory of compatible generality. Now we can
mention three sources for such a theory:
\begin{enumerate}
\item{Theory of individual adaptation of similar individuals to a
similar system of factors;}
\item{Theory of interaction: information
interaction, co-ordination, or deeper integration;}
\item{Theory of collective effects in market dynamics.}
\end{enumerate}

The first approach (supported by biological data) is a sort of
mean-field theory: everybody is adapting to a field of common
factors, and altogether change the state of that system. There are
two types of argumentation here: similarity of factors, or
similarity of adaptation mechanisms (or both):
\begin{itemize}
\item{In the period of crisis the same challenges appear for
most of the market participants, and correlation increases because
they have to answer the same challenge and struggle with the same
factors.} \item{In the period of crisis all participants are under
pressure. The nature of that pressure may be different, but the
mobilization mechanisms are mostly universal. Similar attempts at
adaptation produce correlation as a consequence of crisis.}
\end{itemize}
This theory is focused on the adaptation process, but may be
included into any theory of economical dynamics as adaptation
feedback. We study the adaptation of individuals in the ``mean
field", and consider dynamics of this field as external
conditions.

The interaction theory may be much more rich (and complicated).
For example, it can consider the following effect of behavior in
crisis: there is a lack of information and of known optimal
solutions, therefore, different agents try to find clues to
rational behavior in the behavior of other agents, and the
correlation increases. Coordination in management and in financial
politics is an obvious effect of interaction too, and we can
observe also a deeper integration, which causes fluxes of moneys
and goods.

Collective effects in market dynamics may also generate correlations
and, on the other hand, can interact with correlations which appear
by any specific or nonspecific reasons. For example, high levels of
correlation often lead to the loss of dissipation in dynamics and
may cause instability.

Further in this work, we focus on the theory of individual
adaptation of similar individuals to a similar system of factors.

\section{Theoretical approaches}

\subsection{The ``Energy of Adaptation" and Factors-Resources
Models}

\subsubsection{Factors and Systems}

Let us consider several systems that are under the influence of
several factors $F_1,... F_q$. Each factor has its intensity $f_i$
($i=1,...q$). For convenience, we consider all these factors as
harmful (later, after we introduce fitness $W$,  it will mean that
all partial derivatives are non-positive $\partial W /\partial f_i
\leq 0$, this is a formal definition of ``harm"). This is just a
convention about the choice of axes directions: a wholesome factor
is just a ``minus harmful" factor.

Each system has its adaptation systems, a ``shield" that can
decrease the influence of these factors. In the simplest case, it
means that each system has an available adaptation resource, $R$,
which can be distributed for neutralization of factors: instead of
factor intensities $f_i$ the system is under pressure from factor
values $f_i-a_i r_i$ (where $a_i>0$ is the coefficient of
efficiency of factor $F_i$ neutralization by the adaptation system
and $r_i$ is the share of the adaptation resource assigned for the
neutralization of factor $F_i$, $\sum_i r_i \leq R$). The zero
value $f_i-a_i r_i= 0$ is optimal (the fully compensated factor),
and further compensation is impossible and senseless.

Interaction of  each system with a factor $F_i$ is described by
two quantities: the factor $F_i$ uncompensated pressure
$\psi_i=f_i-a_i r_i$ and the resource assigned to the factor
$F_i$ neutralization. The question about interaction of various
factors is very important, but, first of all, let us study a
one-factor model.

\subsubsection{\label{Selye Model} Selye Model}

Already simple one--factor models support the observed effect of
the correlation increase. In these models, observable properties
of interest $x_k$ $(k=1,...m)$ can be modeled as functions of
factor pressure $\psi$ plus some noise $\epsilon_k$.

Let us consider one-factor systems and linear functions (the
simplest case):
\begin{equation}\label{1factorTension}
x_k=\mu_k + l_k \psi+ \epsilon_k \ ,
\end{equation}
where $\mu_k$ is the mean value of $x_k$ for fully compensated
factor, $l_k$ is a coefficient, $\psi=f-ar_f\geq 0$, and $r_f \leq
R$ is amount of available resource assigned for the factor
neutralization. The values of $\mu_k$ could be considered as
``normal" (in the sense opposite to ``pathology"), and noise
$\epsilon_k$ reflects variability of norm. This is not a dynamic
equation and describes just one action of resource assignment. If we
add time $t$ then a two-dimensional array appears $x_{kt}$.

We can call these models the ``tension--driven models" or even
the ``Selye models" because these models may be extracted from
the Selye concept of adaptation energy \cite{SelyeAEN,SelyeAE1}
(Selye did not use equations, but qualitatively these models
were present in his reasoning).

If systems compensate as much of the factor value, as possible,
then $r_f= \min\{R,f/a\}$, and we can write:
\begin{equation}\label{OneFactor}
 \psi=\left\{\begin{array}{ll}
 &f-a R\ , \ \ {\rm if} \ \ f>aR \ ; \\
 &0 ,\ \  \ {\rm else.}
 \end{array}
 \right.
\end{equation}

The {\it nonlinearity} of the Selye model is in the dependence of
$\psi$ on the factor pressure $f$. Already the simple dependence
(\ref{OneFactor}) gives the phase transition picture. Individual
systems may be different by the value of factor intensity (the
local intensity variability), by the amount of available resource
$R$ and, of course, by the random values $\epsilon_k$. For small
$f$ all $\psi=0$, all systems are in comfort zone and all the
difference between them is in the noise variables $\epsilon_k$. In
this state, the correlations are defined by the correlations in
noise values and are, presumably, low.

With increasing $f$ the separation appears: some systems remain
in the comfort ``condensate" ($\psi=0$), and others already do
not have enough resource for a full compensation of the factor
load and vary in  the value of $\psi$. Two fractions appear, a
lowly correlated condensate with $\psi=0$ and a highly
correlated fraction with different values of $\psi>0$. If $f$
continues to increase, all individuals move to the highly
correlated fraction and the comfort concentrate vanishes.

If the noise of the norm $\epsilon_k$ is independent of $\psi$
then the correlation between different $x_k$ increases
monotonically with $f$. With an increase of the factor intensity
$f$ the dominant eigenvector of the correlation matrix between
$x_k$ becomes more uniform in the coordinates, which tend
asymptotically to $\pm \frac{1}{\sqrt{m}}$.

The correlation between systems also increases (just transpose the
data matrix), and the coordinates of the dominant eigenvector
similarly tend to values $\frac{1}{\sqrt{n}}$ (which are
positive), but this tendency has the character of a ``resource
exhausting wave" which spreads through the systems following the
rule (\ref{OneFactor}).

The observation of Ref.~\cite{Stanley2002} partially supports the
uniformity of the eigenvector that corresponds to the largest
eigenvalue which ``represents the influence of the entire market
that is common to all stocks." Fig.~8d from
Ref.~\cite{Stanley2002} shows that the components of this
eigenvector are positive and ``almost all stocks participate in
the largest eigenvector." Also, in
Ref.~\cite{DrozdComCollNoise2000} it was demonstrated that in the
periods of drawdowns of the global index (DAX) there appears one
strongly dominant eigenvalue for synchronous correlations between
30 companies from DAX. Similar results for 30 British companies
are presented in Figs.~\ref{FTSE_Gt_Dim}, \ref{FTSE_G05_Dim}. In
physiology, we also found these ``maximum integration" effects for
various loads on organisms \cite{Svetlichnaia}. When the pressure
is lower then, instead of one dominant eigenvector which
represents all functional systems of an organism, there appears a
group of eigenvectors with relatively high eigenvalues. Each of
these vectors has significant components for attributes of a
specific group of functional systems, and the intersection of
those groups for different eigenvectors is not large. In addition,
the effect of factor ``disintegration" because of overload was
also observed.

The Selye model describes the first part of the effect (from
comfort to stress), but tells us nothing about the other side of
crisis near the death.

\subsubsection{Mean Field Realization of Selye's Model \label{toy}}

In this Sec. we present a simple toy model that is the mean
field realization of the Selye model. As a harmful factor for
this model we use minus log-return of the FTSE index: the
instant value of factor $f(k)$ at time moment $k$ is
\begin{equation}\label{MeanFTSEret}
f(t)=-\log({\rm FTSE}(t+1)/{\rm FTSE}(t))
\end{equation}
This factor could be considered as the mean field produced by the
all objects together with some outer sources.

The instant values of stocks log-returns of $i$th object
$x_i(k)$ are modeled by the Selye model (\ref{OneFactor}):
\begin{equation}\label{ToyModelEq}
x_i(t)=-l(f(t)- a r_i)H(f(t)- a r_i)+ \epsilon_i (t) \, ,
\end{equation}
where $H$ is the Heaviside step function.

We compare real data and data for two distributions of resource,
Exponential(30) (subscript ``exp") and Uniform(0,2) (subscript
``u"). Random variables $\epsilon_i (t)$ for various $i$ and $t$ are
uniformly distributed i.i.d with zero mean and the variance
var$\epsilon$=0.0035. This is the minimum of the average variance of
the log-return values for thirty companies. The minimum corresponds
to the most ``quiet" state of market (in the sense of value of
variance) in the time period. We calculated the total variance of 30
companies during the time interval used for analysis (04/07/2007 -
25/10/2007), found the minimal value of the variance and divided by
30. To compare results for exponential and uniform distributions we
use the same realization of noise.

The efficiency coefficient $a$ is different for different
distributions: we calibrate it on such a way that for 75\% of
objects the value $ar_i$ is expected to be below $f$ and 25\% are
expected to be above $f$ for the same value of factor $f$: $a_{\rm
exp}/a_{\rm u}\approx 1.88$. The ratio of the coefficients $l_{\rm
exp}/l_{\rm u}$ should have (approximately) inverse value to keep
the expected distances the same for the pairs of objects with $ar_i
< f$. For qualitative reproduction of the crisis we selected $a_{\rm
exp}=0.032$, $a_{\rm u}=0.017$, $l_{\rm exp}=7.3$, $l_{\rm u}=15.5$.

\begin{figure}
\includegraphics[width=150mm]{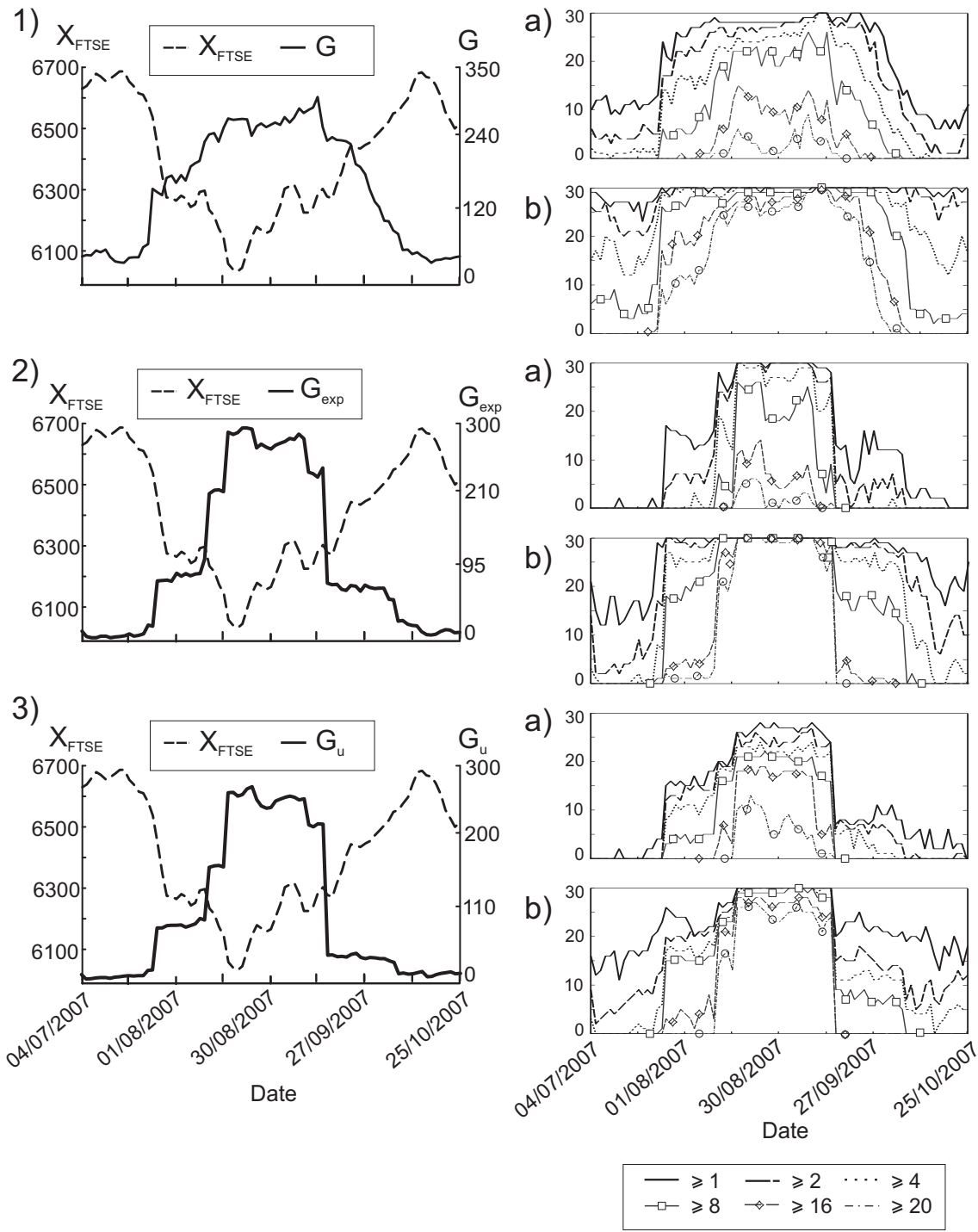}
\caption{\label{ToyModel} The dynamics of indicators of correlation
matrices  for 1) real data, 2) system with exponentially distributed
resources, 3) system with uniformly distributed resources. The
left-hand part represents the general dynamics of $G$, $G_{\rm
exp}$,$G_{\rm u}$ in comparison to the dynamics of FTSE over the
time period 04/07/2007 - 25/10/2007 . The right-hand part shows the
dynamics of changes in correlations between objects over the
interval: a) number of objects that have more than 1, 2, 4, 8, 16 or
20  values of correlations greater than $0.7$, b) number of objects
that have more than 1, 2, 4, 8, 16 or 20  values of correlations
greater than $0.5$.}
\end{figure}

For each system we calculated the correlation coefficients over the
period of 20 days (similar to the analysis made for real data):
$G_{\rm exp}$, $G_{\rm u}$. The right-hand side of the figure
represent the dynamics of changes in correlations between objects.
Plots in Fig.~\ref{ToyModel}.1a show the number of objects in real
data that have more than 1, 2, 4, 8, 16 or 20 values of correlations
greater than $0.7$, plots in Fig.~\ref{ToyModel}1b represent the
number of companies that have more than 1, 2, 4, 8, 16 or 20
correlations greater than $0.5$. Similarly,
Figs.~\ref{ToyModel}.2a,b and~\ref{ToyModel}3.a,b represent the
model results for the exponential (2) and uniform (3) distributions.

The qualitative character of crisis is reproduced, but the
difference from the empirical data is also obvious: the plots for
real data also bell-shaped with fluctuations, but they are wider
than the model curves and fluctuations do not go to zero outside the
crisis period in reality. The simplest improvement of the situation
may be achieved by introduction of correlated noise and fitting. In
the simplest Selye model we assume zero correlations in the comport
zone but in reality the correlations do not decrease to zero.

Amplitude of noise differs for different companies and we can take
its distribution from empirical data. Coefficient $l$ in the basic
Selye model (\ref{1factorTension}) also depends on the company but
in the toy model we take it constant.

One problem exists for all these improvements: they introduce too
many parameters for fitting. Of course, more degrees of freedom
available for fitting give more flexibility in quantitative
approximation of the empirical data. The simplest toy model has two
parameters only.

Another way to improvement is the selection of a better mean field
factor. Now we make just a first choice and selected the negative
log-return of the FTSE index as a mean-field harmful factor. The
serious modification of model could take into account the pressure
of several factors too.

\subsubsection{How to Merge Factors?}

Usually, there are many factors. Formally, for $q$ factors one can
generalize the one--factor tension--driven model
(\ref{1factorTension}) in the form.
\begin{equation}\label{qfactorTension}
x_k=x_k(\psi_1, \psi_2, ... \psi_q)+ \epsilon_k \ .
\end{equation}
In this equation, the {\it compensated} values of factors,
$\psi_i=f_i-a_i r_i$, are used and $\sum_{i=1}^q r_i \leq R$.

Two questions appear immediately: (i) how to find the distribution
of resource, assigned for neutralization of different factors, and
(ii) how to represent the functions $x_k(\psi_1,...\psi_q)$.
Usually, in factor analysis and in physics both, we start from the
assumption of linearity (``in the first approximation"), but this
approximation does not work here properly. In the simplest
reasonable approximation, max-min operations appear instead of
linear operations. This sounds very modern \cite{LitMas} and even
a bit extravagant, but it was discovered many years ago by Justus
von Liebig (1840).  His ``law of the minimum" states that growth
is controlled by the scarcest resource (limiting factor)
\cite{Liebig1}. This concept was originally applied to plant or
crop growth. Many times it was criticized, rejected, and then
returned and demonstrated quantitative agreement with experiments
\cite{Liebig1}, \cite{Liebig2+}, \cite{Liebig3+}. Liebig's Law of
the minimum was extended to more a general conception of factors,
not only for elementary physical description of available chemical
substances and energy. Any environmental factor essential for life
that is below the critical minimum, or that exceeds the maximum
tolerable level could be considered as a limiting one.

The biological generalizations of Liebig's Law were supported by
the biochemical idea of limiting reaction steps (the modern theory
of limiting steps and dominant systems for multiscale reaction
networks is presented in the recent review \cite{GorRadLim}). Some
of the generalizations went quite far from agriculture and
ecology. The law of the minimum was applied to economics
\cite{EcolEcon} and to education, for example \cite{EcolEdu}.

According to Liebig's Law, the tension--driven model is
\begin{equation}\label{LiebqfactorTension}
x_k=\mu_k +l_k \max_{1\leq i\leq q} \{\psi_i\}+ \epsilon_k \ .
\end{equation}
This model seems to be linear, but its nonlinearity is hidden in
dependence of $\psi_i$ on the distribution of factors and the
amount  of the resource available.

\subsubsection{Optimality and Fitness}

Adaptation optimizes the state of the system for a given amount of
the resource available. It may be difficult to find the objective
function that is hidden behind the adaptation process.
Nevertheless, even an assumption about the existence of an
objective function and about its general properties helps in the
analysis of the adaptation process. Assume that adaptation should
maximize an objective function $W$ which depends on the
compensated values of factors, $\psi_i=f_i-a_i r_i$ for the given
amount of available resource:
\begin{equation}\label{Optimality}
\left\{ \begin{array}{l}
 W(f_1-a_1 r_1, f_2-a_2 r_2, ... f_q-a_q r_q) \ \to \ \max \ ; \\
 r_i \geq 0$, $f_i-a_i r_i \geq 0$, $\sum_{i=1}^q r_i \leq R \ .
\end{array} \right.
\end{equation}
The only question is: why can we be sure that adaptation
follows any optimality principle? Existence of optimality is
proven for microevolution processes and ecological succession.
The mathematical backgrounds for the notion of ``natural
selection" in these situations are well established after works
of Haldane (1932) \cite{Haldane} and Gause (1934) \cite{Gause}.
Now this direction with various concepts of {\it fitness} (or
``generalized fitness") optimization is elaborated in many
details (see, for example, review papers
\cite{Bom02,Oechssler02,GorbanSelTth}).

The foundation of optimization is not so clear for such
processes as modifications of phenotype, and for adaptation in
various time scales. The idea of {\it genocopy-phenocopy
interchangeability} was formulated long ago by biologists to
explain many experimental effects: the phenotype modifications
simulate the optimal genotype
(\cite{West-Eberhardgenocopy-phenocopy}, p. 117). The idea of
convergence of genetic and environmental effects was supported
by analysis of genome regulation
\cite{ZuckerkandlConvergGenEnv} (the principle of
concentration-affinity equivalence). The phenotype
modifications produce the same change, as evolution of genotype
does, but faster and in a smaller range of conditions (the
proper evolution can go further, but slower). It is natural to
assume that adaptation in different time scales also follows
the same direction, as evolution and phenotype modifications,
but faster and for smaller changes. This hypothesis could be
supported by many biological data and plausible reasoning. For
social and economical systems the idea of optimization of
individual behavior seems very natural. The selection arguments
may be also valid for such systems.

It seems productive to accept the idea of optimality, and to use
it, as far as this will not contradict the data.

\subsection{Law of the Minimum Paradox \label{Sec:LawMinParad}}

Liebig used the image of a barrel -- now called Liebig's barrel --
to explain his law. Just as the capacity of a barrel with staves
of unequal length is limited by the shortest stave, so a plant's
growth is limited by the nutrient in shortest supply. An
adaptation system acts as a cooper and repairs the shortest stave
to improve the barrel capacity. Indeed, in well-adapted systems
the limiting factor should be compensated as far as this is
possible. It seems obvious because of the very natural idea of
optimality, but arguments of this type in biology should be
considered with care.

Assume that adaptation should maximize a objective function $W$
(\ref{Optimality}) which satisfies Liebig's Law:
\begin{equation}\label{objective}
W=W\left(\max_{1\leq i\leq q} \{f_i-a_i r_i\}\right)\ ; \
\frac{\partial W(x)}{\partial x} \leq 0
\end{equation}
under conditions $r_i \geq 0$, $f_i-a_i r_i \geq 0$, $\sum_{i=1}^q
r_i \leq R$. (Let us recall that $f_i \geq 0$ for all $i$.)

Description of the maximizers of $W$ gives the following theorem
(the proof is a straightforward consequence of Liebig's Law and
monotonicity of $W$).

{\bf Theorem 1}. {\it For any objective function $W$ that
satisfies conditions (\ref{objective}) the optimizers $r_i$ are
defined by the following algorithm.
\begin{enumerate}
\item{Order intensities of factors: $f_{i_1} \geq f_{i_1} \geq ...
f_{i_q}$.}
 \item{Calculate differences $\Delta_j =f_{i_j} -f_{i_{j+1}}$
(take formally $\Delta_0=\Delta_{q+1}=0$).}
 \item{Find such $k$ ($0 \leq
k \leq q$) that
  $$\sum_{j=1}^{k} \left(\sum_{p=1}^j \frac{1}{a_{i_p}}\right) \Delta_j
  \leq R \leq \sum_{j=1}^{k+1} \left(\sum_{p=1}^j
 \frac{1}{a_{i_p}}\right) \Delta_j \ . $$ For $R< \Delta_1$
 we put $k=0$ and if  $R>  \sum_{j=1}^q \left(\sum_{p=1}^j
 \frac{1}{a_{i_p}}\right) \Delta_j$ then we take $k=q$.}
\item{If $k < q$ then the optimal amount of resource $r_{j_l}$ is
\begin{equation}\label{OptimDistr}
r_{j_l}=\left\{\begin{array}{ll}
 &\frac{f_{j_l}}{a_{j_l}} - \frac{1}{a_{j_l}}\left(\sum_{p=1}^k \frac{1}{a_{i_p}}\right)^{-1}
 \left(\sum_{j=1}^k \frac{f_{i_j}}{a_{i_j}}-R\right)\
 , \  \ {\rm if} \ \ l \leq k+1\ ; \\
 &0\ , \ \ \ \ \ {\rm if } \  \ l> k+1\ .
\end{array}
 \right.
\end{equation}
If $k=q$ then $r_i = f_i/a_i$ for all $i$.}
\end{enumerate}
}
\begin{figure} \centering{
\includegraphics[width=100mm]{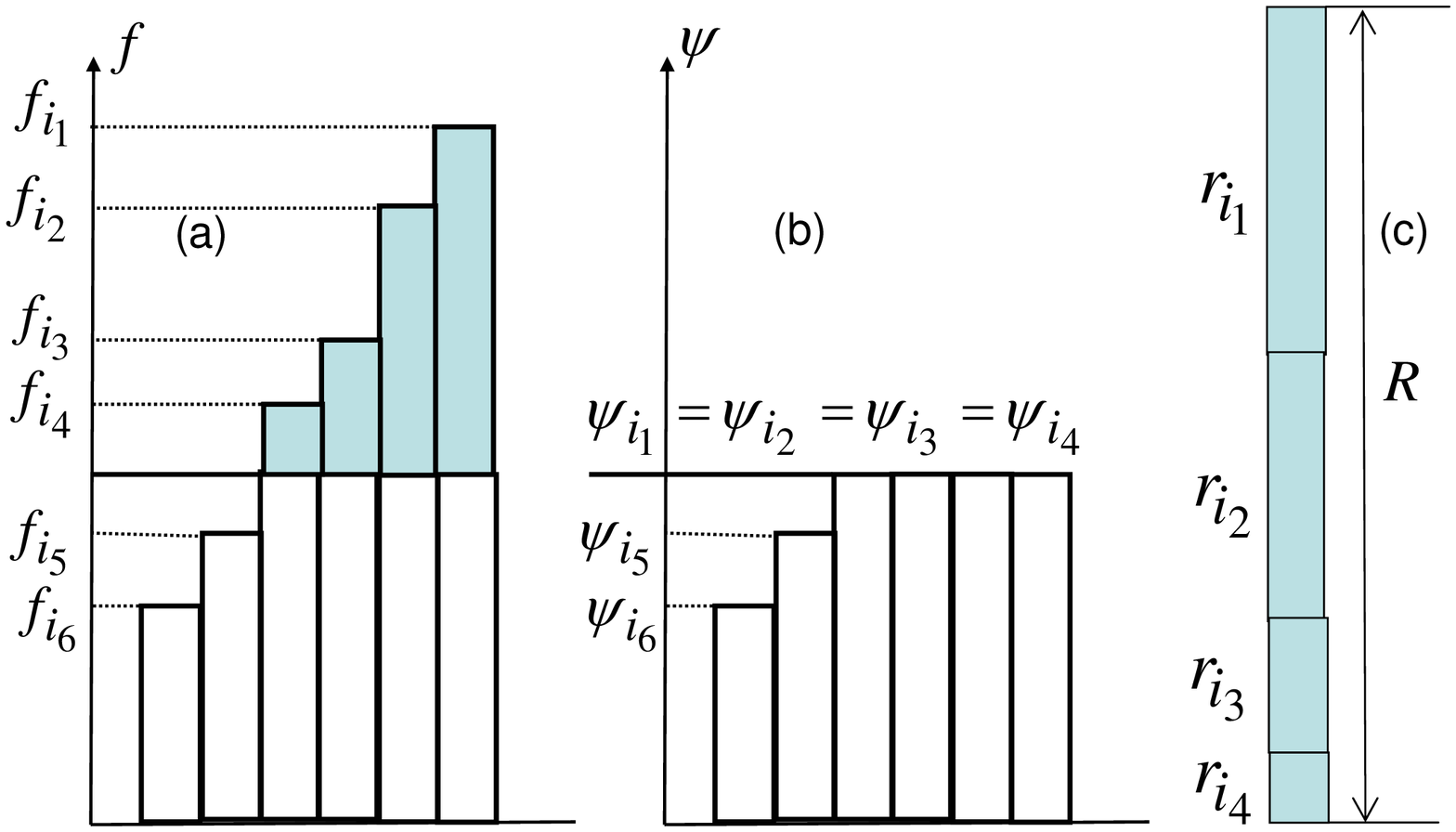}}
\caption{\label{Fig:FactorDistrib} Optimal distribution of resource
for neutralization of factors under Liebig's Law. (a) histogram of
factors intensity (the compensated parts of factors are highlighted,
$k=3$), (b) distribution of tensions $\psi_i$ after adaptation
becomes more uniform, (c) the sum of distributed resources. For
simplicity of the picture, we take here all $a_i=1$.}
\end{figure}

\noindent {\it Proof.} This optimization is illustrated in
Fig.~\ref{Fig:FactorDistrib}. If $R \geq \sum_i
f_{i_j}/a_{i_j}$ then the pressure of all the factors could be
compensated and we can take $r_i=f_i/a_i$. Now, let us assume
that $R < \sum_i f_{i_j}/a_{i_j}$. In this case, the pressure
of some of the factors is not fully compensated. The adaptation
resource is spent for partial compensation of the $k+1$ worst
factors and the remained pressure of them is higher (or equal)
then the pressure of the $k+2$ worst factor:
\begin{equation}\label{extrcond}
\begin{split}
&f_{i_1}-a_{i_1}r_{i_1}=\ldots =f_{i_{k+1}}-a_{i_{k+1}}r_{i_{k+1}}=\psi \geq
f_{i_{k+2}}\, , \; \sum_{j=1}^{k+1} r_{i_j}=R \, , \mbox{or } \\
&\sum_{i=1}^{k+1} \Delta_i -a_{i_1}r_{i_1} = \ldots = \Delta_{k+1} - a_{i_{k+1}}r_{i_{k+1}} =
\psi-f_{i_{k+2}}=\theta_{k+1} \geq 0 \, , \; \sum_{j=1}^{k+1} r_{i_j}=R \, .
\end{split}
\end{equation}
Therefore, for $j=1,\ldots , k+1$ in the optimal distribution
of the resource,
\begin{equation}
r_{i_j}= \frac{1}{a_{i_j}}\left(\sum_{i=j}^{k+1}\Delta_i - \theta_{k+1}\right)\, ,
R=\sum_{j=1}^{k+1}r_{i_j}=\sum_{j=1}^{k+1} \left(\sum_{p=1}^j\frac{1}{a_{i_p}}\right) \Delta_j -
\left(\sum_{j=1}^{k+1}\frac{1}{a_{i_j}}\right) \theta_{k+1}\, , \theta_{k+1}\geq 0\, .
\end{equation}
This gives us the first step in the Theorem 1, the definition
of $k$. Formula (\ref{OptimDistr}) for $r_{i_j}$ follows also
from (\ref{extrcond}): for $j=1, \ldots , k+1$
\begin{equation}
r_{i_j}=\frac{f_{i_j}-\psi}{a_{i_j}}\, , \psi=\left(\sum_{p=1}^{k+1}\frac{1}{a_{i_p}}\right)^{-1}
\left(\sum_{p=1}^{k+1}\frac{f_{i_p}}{a_{i_p}}-R\right)\, . \;\;\;\square
\end{equation}

Hence, if the system satisfies the law of the minimum then the
adaptation process makes the tension produced by different factors
$\psi_i=f_i-ar_i$ (Fig.~\ref{Fig:FactorDistrib}) more uniform.
Thus adaptation decreases the effect of the limiting factor and
hides manifestations of Liebig's Law.

Under the assumption of optimality  (\ref{Optimality}) the {\it law
of the minimum paradox} becomes a theorem: if Liebig's Law is true
then microevolution, ecological succession, phenotype modifications
and adaptation decrease the role of the limiting factors and bring
the tension produced by different factors together.

The cooper starts to repair Liebig's barrel from the shortest stave
and after reparation the staves are more uniform, than they were
before. This cooper may be microevolution, ecological succession,
phenotype modifications, or adaptation. For the ecological
succession this effect (Liebig's  Law leads to its violation by
succession) was described in Ref.~\cite{SemSem}. For adaptation (and
in general settings too) it was demonstrated in
Ref.~\cite{GorSmiCorAd1st}.

The law of the minimum together with the idea of optimality (even
without an explicit form of the objective function) gives us
answers to both question: (i) we now know the optimal distribution
of the resource (\ref{OptimDistr}), assigned for neutralization of
different factors, and (ii) we can choose the function
$x_k(\psi_1,...\psi_q)$ from various model forms, the simplest of
them gives the tension--driven models (\ref{LiebqfactorTension}).

\subsection{Law of the Minimum Inverse Paradox \label{Sec:LawMinInvPar}}

The simplest formal example of ``anti--Liebig's" organization of
interaction between factors gives us the following dependence of
fitness from two factors: $W=-f_1 f_2$: each of the factors is
neutral in the absence of another factor, but together they are
harmful. This is an example of {\it synergy}: the whole is greater
than the sum of its parts. (For our selection of axes direction,
``greater" means ``more harm".) Let us give the formal definition
of the synergistic system of factors for the given fitness
function $W$.

{\bf Definition}. {\it The system of factors $F_1,... F_q$ is
synergistic, if for any two different vectors of their admissible
values $\mathbf{f}=(f_1,... f_q)$ and $\mathbf{g}=(g_1,... g_q)$
($\mathbf{f} \neq \mathbf{g}$)  the value of fitness at the average
point $(\mathbf{f}+\mathbf{g})/2$ is less, than at the best of
points $\mathbf{f}$, $\mathbf{g}$:
\begin{equation}\label{synergy}
W\left(\frac{\mathbf{f}+\mathbf{g}}{2}\right) <
\max\{W(\mathbf{f}),W(\mathbf{g})\}\ .
\end{equation}}

Liebig's systems of factors violate the synergy inequality
(\ref{synergy}): if at points $\mathbf{f}$, $\mathbf{g}$ with the
same values of fitness $W(\mathbf{f})=W(\mathbf{g})$ different
factors are limiting, then at the average point the value of both
these factors are smaller, and the harm of the limiting factor at
that point is less, than at both points $\mathbf{f}$,
$\mathbf{g}$, i.e. the fitness at the average point is larger.

The fitness function $W$ for synergistic systems has a property that
makes the solution of optimization problems much simpler. This
proposition follows from the definition of convexity and standard
facts about convex sets (see, for example, \cite{Rockafellar})

{\bf Proposition 1}. {\it The synergy inequality (\ref{synergy})
holds if and only if all the sublevel sets $\{\mathbf{f} \ | \
W(\mathbf{f}) \leq \alpha \}$ are strictly convex.}$\square$

(The fitness itself may be a non-convex function.)

This proposition immediately implies that the synergy inequality is
invariant with respect to increasing monotonic transformations of
$W$. This invariance with respect to nonlinear change of scale is
very important, because usually we don't know the values of function
$W$.

{\bf Proposition 2}. {\it If the synergy inequality (\ref{synergy})
holds for a function $W$, then it holds for a function
$W_{\theta}=\theta (W)$, where $\theta(x)$ is an arbitrary strictly
monotonic function of one variable.}$\square$

Already this property allows us to study the problem about optimal
distribution of the adaptation resource without further knowledge
about the fitness function.

Assume that adaptation should maximize an objective function
$W(f_1-r_1, ... f_q-r_q)$  (\ref{Optimality}) which satisfies the
synergy inequality (\ref{synergy}) under conditions $r_i \geq 0$,
$f_i-a_i r_i \geq 0$, $\sum_{i=1}^q r_i \leq R$. (Let us remind that
$f_i \geq 0$ for all $i$.) Following our previous convention about
axes directions all factors are harmful and $W$ is monotonically
decreasing function
$$\frac{\partial W(f_1,... f_q)}{\partial f_i} < 0 \ .$$ We need
also a technical assumption that $W$ is defined on a convex set in
$\mathbb{R}^q_+$ and if it is defined for a nonnegative point
$\mathbf{f}$, then it is also defined at any nonnegative point
$\mathbf{g} \leq \mathbf{f}$ (this inequality means that $g_i \leq
f_i$  for all $i=1,... q$).

The set of possible maximizers is finite. For every group of
factors $F_{i_1},... F_{i_{j+1}}$, ($1 \leq j+1 < q$) with the
property
\begin{equation}\label{compensation}
\sum_{k=1}^j \frac{f_{i_k}}{a_{i_k}} < R \leq \sum_{k=1}^{j+1}
\frac{f_{i_k}}{a_{i_k}}
\end{equation}
we find a distribution of resource $\mathbf{r}_{\{{i_1},...
{i_{j+1}}\}}=(r_{i_1},... r_{i_{j+1}})$:
\begin{equation}\label{antiLiebigDistr}
r_{i_k}=\frac{f_{i_k}}{a_{i_k}} \ \ (k=1,... j)\ , \ \
r_{i_{j+1}}=R-\sum_{k=1}^j \frac{f_{i_k}}{a_{i_k}}\ , \ \ r_i=0 \
\ {\rm for }\ \ i \notin \{{i_1},... {i_{j+1}}\} \ .
\end{equation}
For $j=0$, Eq.~(\ref{compensation}) gives $0 < R \leq
f_{i_1}/a_{i_1}$ and there exists only one nonzero component in
the distribution (\ref{antiLiebigDistr}),
$r_{i_{1}}=R/a_{i_{1}}$.

We get the following theorem as an application of standard results
about extreme points of convex sets \cite{Rockafellar}.

{\bf Theorem 2}. {\it Any maximizer for $W(f_1-r_1, ... f_q-r_q)$
under given conditions has the form $\mathbf{r}_{\{{i_1},...
{i_{j+1}}\}}$ (\ref{antiLiebigDistr}).}$\square$

\begin{figure} \centering{
\includegraphics[width=100mm]{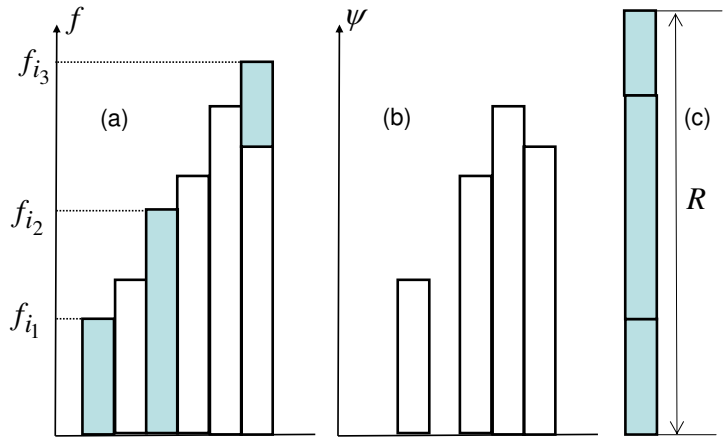}}
\caption{\label{Fig:FactorDistribSyn} Typical optimal distribution
of resource for neutralization of synergistic factors. (a) Factors
intensity (the compensated parts of factors are highlighted, $j=2$),
(b) distribution of tensions $\psi_i$ after adaptation becomes less
uniform (compare to Fig.~\ref{Fig:FactorDistrib}), (c) the sum of
distributed resources. For simplicity of the picture, we take here
all $a_i=1$.}
\end{figure}

If the initial distribution of factors intensities,
$\mathbf{f}=(f_1,... f_q)$, is almost uniform and all factors
are significant then, after adaptation, the distribution of
effective tensions, $\mathbf{\psi}=(\psi_1,... \psi_q)$
($\psi_i=f_i-a_i r_i$), is less uniform. Following Theorem 2,
some of the factors may be completely neutralized and one
additional factor may be neutralized partially. This situation
is opposite to adaptation due to Liebig's system of factors,
where the amount of significant factors increases and the
distribution of tensions becomes more uniform because of
adaptation. For Liebig's system, adaptation transforms a
low-dimensional picture (one limiting factor) into a
high-dimensional one, and we expect the well-adapted systems
have less correlations than in stress. For synergistic systems,
adaptation transforms a high-dimensional picture into a
low-dimensional one (less factors), and our expectations are
inverse: we expect the well-adapted systems have more
correlations than in stress (this situation is illustrated in
Fig.~\ref{Fig:FactorDistribSyn}; compare to
Fig.~\ref{Fig:FactorDistrib}). We call this property of
adaptation to synergistic system of factors the {\it law of the
minimum inverse paradox}.

Fitness by itself is a theoretical construction based on the
average reproduction coefficient (instant fitness). It is
impossible to measure this quantity in time intervals that are
much shorter than life length. Hence, to understand which system
of factors we deal with, Liebig's or a synergistic one, we have to
compare the theoretical consequences of their properties. First of
all, we can measure the results of adaptation, and use properties
of the optimal adaptation in ensembles of systems for analysis
(Fig.~\ref{Fig:FactorDistrib}, Fig.~\ref{Fig:FactorDistribSyn}).

There is some evidence about the existence of synergistic systems
of factors. For example, the postsurgical rehabilitation of people
suffering lung cancer of the III and IV clinical groups was
studied \cite{Mansurov}. Dynamics of variance and correlations for
them have directions which are unusual for Liebig's systems:
increase of the correlation corresponds to decrease of the
variance. Moreover, analysis of the maxima and minima of
correlations and mortality demonstrates that in this case an
increase of correlations corresponds to decrease of stress. Hence,
in Ref.~\cite{Mansurov} the hypothesis was suggested that in this
case some factors superlinearly increase the harmfulness of other
factors, and this is an example of a synergistic system of
factors. Thus, the law of the minimum inverse paradox may give us
a clue to the effect (Fig.~\ref{Fig:effect}) near the fatal
outcomes.

\section{Discussion}

\subsection{Dynamics of the Correlations in Crisis}

We study a universal effect in ensembles of similar systems under
load of similar factors: in crisis, typically, correlation
increases, and, at the same time, variance (and volatility)
increases too. This effect is demonstrated for humans, mice, trees,
grassy plants, and financial time series. It is represented as the
left transition in Fig.~\ref{Fig:effect}, the transition from
comfort to crisis. Already a system of simple models of adaptation
to one factor (we call it the {\it Selye model}) gives a qualitative
explanation of the effect.

For interaction of several factors two basic types of
organization are considered: Liebig's systems and synergistic
systems of factors. The adaptation process (as well as
phenomodification, ecological succession, or microevolution)
acts differently onto these systems of factors and makes
Liebig's systems more uniform (instead of systems with limiting
factor) and synergistic systems less uniform. These theorems
give us two paradoxes which explain differences observed
between artificial (less adapted) systems and natural
(well-adapted) systems.

Empirically, we expect the appearance of synergistic systems in
extremely difficult conditions, when factors appear that
superlinearly amplify the harm from other factors. This means that
after the crisis achieves its bottom, it can develop into two
directions: recovering (both correlations and variance decrease) or
fatal catastrophe (correlations decrease, but variance not). The
transition to fatal outcome is represented as the right transition
in Fig.~\ref{Fig:effect}. Some clinical data support these
expectations.

\subsection{Correlations Between the Thirty Largest FTSE Companies}

The case study of the thirty largest companies from British stock
market for the period  2006--2008 supports the hypothesis about
increasing of the correlations in crisis. It is also demonstrated
that the correlation in time (between daily data) also has
diagnostic power (as well as the correlation between companies has)
and connections between days (Figs.~\ref{Time_graph_int1},
\ref{Time_graph_int3}) may clearly indicate and, sometimes, predict
the chronology of the crisis. This approach (use of two time moments
instead of the time window) allows to overcome a smearing effect
caused by usage of time windows (about this problem see
\cite{Onella1,Onella2}).

The principal component analysis demonstrates that the largest
eigenvalues of the correlation matrices increase in crisis and
under environmental pressure (before the inverse effect ``on the
other side of crisis" appears). Different methods for selection of
significant principal components, Kaiser's rule, random matrix
approach and the broken stick model, give similar results in a
case study. Kaiser's rule gives more principal components than two
other methods and the higher sensitivity of the indicator
Dim$_{\rm K}$ causes some difficulties in interpretation. The
random matrix estimates select too small amount of components, and
the indicator Dim$_MP$ seems not sensitive enough. In our case
study the best balance between sensitivity and stability gives the
dimension, estimated by the broken stick model Dim$_{\rm BS}$.

\subsection{Choice of Coordinates and the Problem of Invariance}

All indicators of the level of correlations are non-invariant
with respect to transformations of  coordinates. For example,
rotation to the principal axis annuls all the correlations.
Dynamics of variance also depends on nonlinear transformations
of scales. Dimensionless variance of logarithms (or ``relative
variance") often demonstrates more stable behavior especially
when changes of mean values are large.

The observed effect depends on the choice of attributes.
Nevertheless, many researchers observed it without a special
choice of coordinate system. What does it mean? We can propose a
hypothesis: the effect may be so strong that it is almost
improbable to select a coordinate system where it vanishes. For
example, if one accepts the Selye model (\ref{1factorTension}),
(\ref{OneFactor}) then observability of the effect means that for
typical nonzero values of $\psi$ in crisis
\begin{equation}\label{lrgeeffect}
l_k^2 \psi^2> {\rm var}(\epsilon_k)
\end{equation}
for more than one value of $k$, where var stands for variance of
the noise component (this is sufficient for increase of the
correlations). If $$\psi^2 \sum_k l_k^2 \gg \sum_k {\rm
var}(\epsilon_k)$$ and the set of allowable transformations of
coordinates is bounded (together with the set of inverse
transformations), then the probability to select randomly a
coordinate system which violates condition (\ref{lrgeeffect}) is
small (for reasonable definitions of this probability and of the
relation $\gg$). On another hand, the choice of attributes is
never random, and one can look for the reason of so wide
observability of the effect in our (human) ways to construct the
attribute systems.

\subsection{Two Programs for Further Research}

First of all, the system of  simple models of adaptation should be
fitted to various data, both economical and biophysical. Classical
econometrics \cite{Judge} already deals with hidden factors, now we
have just to fit a special nonlinear model of adaptation to these
factors.

Another possible direction is the development of dynamical models
of adaptation. In the present form the model of an adaptation
describes a single action, distribution of adaptation resource. We
avoid any kinetic modeling. Nevertheless, adaptation is a process
in time. We have to create a system of models with a minimal
number of parameters.

Models of individual adaptation could explain effects caused by
external factors or individual internal factors. They can be also
used with the mean-field models when the interaction between
systems is presented as an additional factor. The models of
interaction need additional hypotheses and data. In this paper, we
do not discuss such models, but in principle they may be
necessary, because crisis may be caused not by purely external
factors but by combination of external factors, individual
internal dynamics and interaction between systems.

{\bf Acknowledgements}. We are very grateful to many people for
21 years of collaboration, to our first co-author
\cite{GorSmiCorAd1st} V.T. Manchuk, and to A.G. Abanov, G.F.
Bulygin, R.A. Belousova, R.G. Khlebopros, G.B. Kofman, A.S.
Mansurov, T.P. Mansurova, L.S. Mikitin,  A.V. Pershin, L.I.
Pokidysheva, M.G Polonskaya, L.D. Ponomarenko, V.N.
Razzhevaikin, K.R. Sedov, S.M. Semenov, E.N. Shalamova, S.Y.
Skobeleva, and G.N. Svetlichnaia. Many physiological data were
collected from Institute for Medical Problems of Northern
Regions\footnote{State Research Institute for Medical Problems
of Northern Regions, Siberian Branch of Russian (USSR) Academy
of Medical Sciences (Krasnoyarsk).}. We also thank the editor
and the anonymous referees of Physica A for careful reading and
the fruitful criticism.


{\small

\begin{thebibliography}{99}

\bibitem{GorSmiCorAd1st}A.N. Gorban, V.T. Manchuk, E.V.
Petushkova (Smirnova), Dynamics of physiological paramethers
correlations and the ecological-evolutionary principle of
polyfactoriality, Problemy Ekologicheskogo Monitoringa i
Modelirovaniya Ekosistem [The Problems of Ecological Monitoring and
Ecosystem Modelling], Vol. 10. Gidrometeoizdat, Leningrad, 1987, pp.
187--198.

\bibitem{Sedov}K.R. Sedov, A.N. Gorban', E.V. Petushkova (Smirnova), V.T. Manchuk, E.N. Shalamova,
Correlation adaptometry as a method of screening of the
population, Vestn. Akad. Med. Nauk SSSR. 1988;(10):69--75. PMID:
3223045

\bibitem{Mansurov}A.S. Mansurov, T.P. Mansurova, E.V. Smirnova, L.S. Mikitin,
A.V. Pershin, How do correlations between physiological parameters
depend on the influence of different systems of stress factors?
Global \& Regional Ecological Problems, R.G. Khlebopros (Ed.),
Krasnoyarsk State Technical University Publ., 1994, JSBN
5-230-08348-4, pp. 499-516.

\bibitem{Strygina}Strygina S.O., Dement'ev S.N., Uskov V.M., Chernyshova G.I.,
Dynamics of the system of correlations between physiological
parameters in patients after myocardial infarction, in:
Mathematics, Computer, Education, Proceedings of conference, Issue
7, Moscow, 2000, pp. 685--689.


\bibitem{Pokidysheva}L.I. Pokidysheva, R.A. Belousova, E.V. Smirnova, Method of adaptometry in
the evaluation of gastric secretory function in children under
conditions of the North, Vestn. Ross. Akad. Med. Nauk.
1996;(5):42--5. PMID: 8924826

\bibitem{Svetlichnaia}G.N. Svetlichnaia, E.V. Smirnova, L.I. Pokidysheva, Correlational
adaptometry as a method for evaluating cardiovascular and
respiratory interaction. Fiziol. Cheloveka 23(3) (1997) 58--62.
PMID: 9264951

\bibitem{RazzhevaikinObese2007}A.V. Vasil'ev, G.Iu. Mal'tsev, Iu.V. Khrushcheva, V.N. Razzhevaikin,
M.I. Shpitonkov. Applying method of correlation adaptometry for
evaluating of treatment efficiency of obese patients, Vopr. Pitan.
76(2) (2007), 36--38. PMID: 17561653

\bibitem{mice}L.D. Ponomarenko, E.V. Smirnova, Dynamical characteristics of blood system
in mice with phenilhydrazin anemiya, Proceeding of 9th International
Symposium ``Reconstruction of homeostasis", Krasnoyarsk, Russia,
March 15-20, 1998, vol. 1, 42--45.

\bibitem{RazzhevaikinTrava1996}I.V. Karmanova, V.N. Razzhevaikin, M.I. Shpitonkov,
Application of correlation adaptometry for estimating a response of
herbaceous species to stress loadings, Doklady Botanical Sciences,
Vols. 346--348, 1996 January–June, 4--7. [Translated from Doklady
Akademii Nauk SSSR, 346, 1996.]

\bibitem{KofmantREES}P.G. Shumeiko, V.I. Osipov, G.B. Kofman, Early detection of
industrial emission impact on Scots Pine needles by composition of
phenolic compounds, Global \& Regional Ecological Problems, R.G.
Khlebopros (Ed.), Krasnoyarsk State Technical University Publ.,
1994, JSBN 5-230-08348-4, 536--543.

\bibitem{LonginCorrNonconst1995}F. Longin, B. Solnik, Is the correlation in
international equity returns constant: 1960-1990? J. International
Money and Finance 14 (1) (1995), 3--26.

\bibitem{MericEuMarket1987}I. Meric, G. Meric, Co-Movements of European Equity Markets
before and after the 1987 Crash, Multinational Finance J.,  1 (2)
(1997), 137--152.

\bibitem{DrozdComCollNoise2000}S. Dro\.{z}d\.{z}, F. Gr\"ummer , A.Z. G\'orski, F. Ruf, J. Speth,
Dynamics of competition between collectivity and noise in the stock
market, Physica A 287 (2000) 440--449.

\bibitem{Gower1966}J.C. Gower,  (1966) Some distance properties of latent root and
vector methods used in multivariate analysis, Biometrika, 53,
325--338.

\bibitem{Stanley2000}R.N. Mantegna,  H.E. Stanley, An Introduction to Econophysics:
Correlations and Complexity in Finance, Cambridge: Cambridge
University Press, (1999).

\bibitem{Mantegna1999}R.N. Mantegna, Hierarchical structure in financial markets,
Eur. Phys. J. B 11 (1) (1999), 193--197.

\bibitem{ChakrabortiMarketCorr2003}J.-P. Onnela, A. Chakraborti, K. Kaski, J. Kert\'esz, A. Kanto,
Dynamics of market correlations: Taxonomy and portfolio analysis,
Phys. Rev. E 68 (2003), 056110.

\bibitem{Stanley1999}P. Gopikrishnan, B. Rosenow, L.A.N. Amaral, H.E. Stanley
Universal and Nonuniversal Properties of Cross Correlations in
Financial Time Series, Phys. Rev. Lett. 83 (1999), 1471--1474.

\bibitem{Stanley2002}V. Plerou, P. Gopikrishnan, B. Rosenow, L.A.N. Amaral, T. Guhr,
H.E. Stanley, Random matrix approach to cross correlations in
financial data, Phys. Rev. E 65 (2002), 066126.

\bibitem{Potters2005}M. Potters, J.P. Bouchaud,  L. Laloux,  Financial applications
of random matrix theory: old laces and new pieces, Acta Phys. Pol. B
36 (9) (2005), 2767--2784.

\bibitem{Wishart}J. Wishart, The generalised product moment distribution
in samples from a normal multivariate population, Biometrika 20A
(1-2) (1928), 32-–52.

\bibitem{Heimo2008}T. Heimo, G. Tibely, J. Saramaki, K. Kaski,
    J. Kertesz, Spectral methods and cluster structure in
    correlation-based networks, Physica A 387 (23) (2008),
    5930--5945.


\bibitem{SadikCrosCorFin2007}S. \c{C}ukur, M. Eryi\~git, R. Eryi\~gt, Cross
correlations in an emerging market financial data, Physica A 376
(2007) 555--564.

\bibitem{MatesanzDependenceCriz2008}D. Matesanz, G.J. Ortega, Network analysis of
exchange data: Interdependence drives crisis contagion, MPRA Paper
No. 7720, posted 12 March 2008;  e-print:
http://mpra.ub.uni-muenchen.de/7720/

\bibitem{Smith2009}R. Smith, The Spread of the Credit Crisis: View from a
Stock Correlation Network (February 23, 2009);  e-print
http://ssrn.com/abstract=1325803

\bibitem{EliassonCrizDefin2001}A.C. Eliasson, C. Kreuter, On currency crisis: A continuous
crisis definition (Deutsche Bank Research Quantitative Analysis
Report), Conference paper,  X International "Tor Vergata" Conference
on Banking and Finance, December 2001.

\bibitem{GorSmiTyuArX}A.N. Gorban, E.V. Smirnova, T.A. Tyukina,
e-print: arXiv:0905.0129v2, 2009.

\bibitem{SelyeAEN}H. Selye, Adaptation Energy, Nature 141 (3577) (21 May
1938), 926.

\bibitem{SelyeAE1}H. Selye, Experimental evidence supporting the conception of
``adaptation energy", Am. J. Physiol. 123 (1938), 758--765.

\bibitem{GP_AE1952}B. Goldstone, The general practitioner and the general adaptation
syndrome, S. Afr. Med. J.  26 (1952), 88--92, 106--109  PMID:
14901129, 14913266.

\bibitem{AEencicl}R. McCarty, K. Pasak, Alarm phase and general adaptation
syndrome, in: Encyclopedia of Stress, George Fink (ed.), Vol. 1,
Academic Press, 2000,  126--130.

\bibitem{BreznitzAEappl}S. Breznitz (Ed.), The Denial of Stress, New York: International
Universities Press, Inc., 1983.

\bibitem{SchkadeOccAdAE2003}J.K. Schkade, S. Schultz,  Occupational Adaptation in
Perspectives, Ch. 7 in: Perspectives in Human Occupation:
Participation in Life, By Paula Kramer, Jim Hinojosa, Charlotte
Brasic Royeen (eds), Lippincott Williams \& Wilkins, Baltimore, MD,
2003, 181--221.

\bibitem{Haldane}J.B.S. Haldane, The Causes of Evolution, Princeton Science Library,
Princeton University Press,  1990.

\bibitem{Gause}G.F. Gause, The struggle for existence, Williams and Wilkins, Baltimore, 1934.
Online: http://www.ggause.com/Contgau.htm.

\bibitem{Bom02}I.M. Bomze, {Regularity vs. degeneracy in dynamics, games, and
optimization: a unified approach to different aspects.} SIAM Review,
2002, Vol. 44, 394-414.

\bibitem{Oechssler02}J.~Oechssler,  F.~Riedel, {On the Dynamic Foundation of
Evolutionary Stability in Continuous Models.} J. Economic Theory,
2002, Vol. 107, 223-252.

\bibitem{GorbanSelTth}A.N. Gorban, Selection Theorem for Systems with Inheritance, Math.
Model. Nat. Phenom., 2 (4) (2007), 1--45; e-print:
cond-mat/0405451

\bibitem{ZuckerkandlConvergGenEnv}E. Zuckerkandl,  R. Villet,
    Concentration-affinity equivalence in gene regulation:
    Convergence of genetic and environmental effects, PNAS
    U.S.A., 85 (1988), 4784--4788.

\bibitem{West-Eberhardgenocopy-phenocopy}M.J. West-Eberhard, Developmental Plasticity and Evolution,
Oxford University Press US, 2003.

\bibitem{VarVolLillo2000}F. Lillo, R.N. Mantegna, Variety and
    volatility in financial markets, Phys. Rev. E 62 (2000),
    6126; e-print cond-mat/0002438.

\bibitem{Whittaker1990}J. Whittaker, Graphical Models in Applied Multivariate Statistics.
Wiley, Chichester, 1990.

\bibitem{Brillinger1996}D.R. Brillinger, Remarks concerning graphical models for time
series and point processes. Revista de Econometria, 16 (1996),
1--23.

\bibitem{Fried2003}R. Fried, V. Didelez, V. Lanius, Partial correlation graphs and
dynamic latent variables for physiological time series, in: Baier,
Daniel (ed.) et al., Innovations in classification, data science,
and information systems. Proceedings of the 27th annual conference
of the Gesellschaft fur Klassifikation e. V., Cottbus, Germany,
March 12?14, 2003. Springer, Berlin, 2005, 259--266.

\bibitem{Verma2005}V. Verma, N. Gagvani, Visualizing Intelligence Information Using
Correlation Graphs, Proc. SPIE, Vol. 5812(2005), 271--282.

\bibitem{Huynh2006}X.-H. Huynh, F. Guillet and H. Briand, Evaluating Interestingness
Measures with Linear Correlation Graph, in: Advances in Applied
Artificial Intelligence, Lecture Notes in Computer Science, 4031,
Springer Berlin - Heidelberg, 2006, p. 312--321

\bibitem{Onella1}J.-P. Onnela, A. Chakraborti, K. Kaski and J. Kert\'esz, Dynamic asset trees and the Black Monday,
Physica A 324 (2003), 247--252.

\bibitem{Onella2}J.-P. Onnela, K. Kaski and J. Kert\'esz, Clustering and information in correlation based financial
networks, Eur. Phys. J. B 38 (2004), 353--362.


\bibitem{CangelosiBrockStick}R. Cangelosi, A. Goriely, Component retention in principal component
analysis with application to cDNA microarray data, Biology Direct
(2007), 2:2. Online: http://www.biology-direct.com/content/2/1/2

\bibitem{Sengupta1999}A.M. Sengupta, P.P. Mitra, Distributions of singular values
for some random matrices, Phys. Rev. E 60 (1999), 3389--3392.

\bibitem{Bul1Limf}G.V. Bulygin, A.S. Mansurov, T.P. Mansurova, E.V. Smirnova, Dynamics
of parameters of human metabolic system during the short-term
adaptation, Institute of Biophysics, Russian Academy of Sciences,
Preprint 180B, 1992.

\bibitem{Bul2Limf}G.V. Bulygin, A.S. Mansurov, T.P. Mansurova, A.A. Mashanov, E.V.
Smirnova, Impact of health on the ecological stress dynamics.
Institute of Biophysics, Russian Academy of Sciences, Preprint 185B,
Krasnoyarsk, 1992.

\bibitem{MansurOnco}A.S. Mansurov, T.P. Mansurova, E.V. Smirnova, L.S. Mikitin, A.V.
Pershin, Human adaptation under influence of synergic system of
factors (treatment of oncological patients after operation),
Institute of Biophysics Russian Academy of Sciences, Preprint 212B
Krasnoyarsk, 1995.

\bibitem{Goetzmann2004}W.N. Goetzmann, L. Li, K.G. Rouwenhorst,
Long--Term Global Market Correlations (October 7, 2004). Yale ICF
Working Paper No. 08-04;    e-print:
http://ssrn.com/abstract=288421

\bibitem{LitMas}G.L.~Litvinov, V.P.~Maslov (Eds.), Idempotent mathematics and
mathematical physics, Contemporary Mathematics, AMS, Providence, RI,
2005.

\bibitem{Liebig1}F. Salisbury, Plant Physiology,
1992. Plant physiology (4th ed.). Wadsworth Belmont, CA

\bibitem{Liebig2+}Q. Paris. The Return of von Liebig's ``Law of the Minimum", Agron.
J., 84 (1992), 1040--1046

\bibitem{Liebig3+}B.S. Cade, J.W. Terrell, R.L. Schroeder, Estimating effects of
limiting factors with regression quantiles, Ecology 80 (1) (1999),
311--323.

\bibitem{GorRadLim}A.N. Gorban, O. Radulescu,
Dynamic and Static Limitation in Multiscale Reaction Networks,
Revisited, Advances in Chemical Engineering 34 (2008), 103--173.

\bibitem{EcolEcon}H.E. Daly, Population and Economics -- A
    Bioeconomic Analysis,
 Population and Environment 12 (3) (1991), 257--263.

\bibitem{EcolEdu}M.Y. Ozden, Law of the Minimum in Learning. Educational
Technology \& Society 7 (3)   (2004), 5--8.

\bibitem{SemSem}F.N.~Semevsky,  S.M.~Semenov. Mathematical modeling of ecological
processes. Gidrometeoizdat, Leningrad,  1982 [in Russian].

\bibitem{Rockafellar}R.T. Rockafellar, Convex analysis, Princeton University Press,
Princeton, NJ, 1970. Reprint: 1997.

\bibitem{Markose}S.M. Markose,  Computability and Evolutionary Complexity:
Markets as Complex Adaptive Systems (CAS). Economic J. 115 (504)
(2005), F159--F192;  e-print: http://ssrn.com/abstract=745578

\bibitem{Judge}G.G. Judge, W.E. Griffiths, R.C.
Hill, H. L\"utkepohl, T.-C. Lee,  The Theory and Practice of
Econometrics, Wiley Series in Probability and Statistics, \#~49 (2nd
ed.), Wiley, New York 1985.

\end{thebibliography}
                                                                                   }
\end{document}